\documentclass[aps,pr,onecolumn,superscriptaddress]{revtex4}
\usepackage{amsmath}
\usepackage{amssymb}
\usepackage{tabularx}
\usepackage{amsthm}
\usepackage{amsfonts}

\usepackage{enumerate}

\usepackage{makecell}
\usepackage{multirow}
\usepackage{overpic}
\usepackage[english]{babel}
\usepackage[most]{tcolorbox}
\usepackage{listings}

\newif\ifmarked
\markedtrue   

\usepackage{xcolor}
\usepackage[normalem]{ulem}

\ifmarked
  
  \newcommand{\del}[1]{\textcolor{red}{\sout{#1}}}
  
\else
  
  \newcommand{\del}[1]{}
  
\fi

\newtcolorbox{cmdbox}{
  enhanced,
  colback=white,       
  boxrule=0.4pt,       
  sharp corners,
  fontupper=\ttfamily,
  left=2mm,right=2mm,top=1mm,bottom=1mm,
  overlay={%
    \draw (frame.south west) -- (frame.south east);
    \draw ([yshift=-2.5pt,xshift=2.5pt]frame.south west) -- ([yshift=-2.5pt,xshift=2.5pt]frame.south east);
    \draw ([yshift=-2.5pt,xshift=2.5pt]frame.north east) -- ([yshift=-2.5pt,xshift=2.5pt]frame.south east);
  },
}

\newcolumntype{L}[1]{>{\raggedright\arraybackslash}p{#1}} 
\newcolumntype{C}[1]{>{\centering\arraybackslash}p{#1}}

\newcommand{\beginsupplement}{%
        \setcounter{table}{0}
        \renewcommand{\thetable}{S\arabic{table}}%
        \setcounter{figure}{0}
        \renewcommand{\thefigure}{S\arabic{figure}}%
     }

\newcommand{\braket}[2]{\left\langle #1 | #2 \right\rangle}

\newcommand{\ket}[1]{\left|#1\right\rangle}

\newcommand{\up}{\uparrow}

\newcommand{\bk}{{\bf k}}
\newcommand{\bv}{{\bf v}}
\newcommand{\br}{{\bf r}}
\newcommand{\bt}{{\bf t}}
\newcommand{\bg}{{\bf g}}
\newcommand{\bG}{{\bf G}}
\newcommand{\bL}{{\bf L}}

\def\ie{{\it i.e.},\ }
\def\eg{{\it e.g.}\ }
\def\ea{{\it et al.}}

\usepackage{hyperref}
\hypersetup{
 pdfnewwindow=true, colorlinks=true,
 linkcolor=blue, anchorcolor=blue,
 citecolor=blue, filecolor=blue,
 menucolor=blue, urlcolor=blue}

\begin{document}

\tolerance 10000

\newcommand{\vk}{{\bf k}}

\title{{\ttfamily IRSSG}: An Open-Source Software Package for Spin Space Groups}

	\author{Sheng Zhang}
    \thanks{These authors contributed equally to this work.}
	\affiliation{Beijing National Laboratory for Condensed Matter Physics,
		and Institute of Physics, Chinese Academy of Sciences, Beijing 100190, China}
	\affiliation{University of Chinese Academy of Sciences, Beijing 100049, China}

	\author{Ziyin Song}
    \thanks{These authors contributed equally to this work.}
	\affiliation{Beijing National Laboratory for Condensed Matter Physics,
		and Institute of Physics, Chinese Academy of Sciences, Beijing 100190, China}
	\affiliation{University of Chinese Academy of Sciences, Beijing 100049, China}

	\author{Zhong Fang}
	\affiliation{Beijing National Laboratory for Condensed Matter Physics,
		and Institute of Physics, Chinese Academy of Sciences, Beijing 100190, China}
	\affiliation{University of Chinese Academy of Sciences, Beijing 100049, China}
    
	\author{Hongming Weng}
	\affiliation{Beijing National Laboratory for Condensed Matter Physics,
		and Institute of Physics, Chinese Academy of Sciences, Beijing 100190, China}
	\affiliation{Condensed Matter Physics Data Center, Chinese Academy of Sciences, Beijing 100190, China}
    
    \author{Zhijun Wang}
	\email{zjwang11@hotmail.com}
    \altaffiliation{\\$~$ The codes are available in the public repository: \url{https://github.com/zjwang11/IRSSG/}.}
	\affiliation{Beijing National Laboratory for Condensed Matter Physics,
		and Institute of Physics, Chinese Academy of Sciences, Beijing 100190, China}
	\affiliation{Condensed Matter Physics Data Center, Chinese Academy of Sciences, Beijing 100190, China}
    \date{\today}
	
\begin{abstract}
We present an open-source software package {\ttfamily IRSSG} for investigating magnetic systems with spin space groups (SSGs). The package works within the density functional theory (DFT) framework and requires wavefunctions from DFT codes, such as VASP, Quantum ESPRESSO, as well as any other code that has an interface to Wannier90. We introduce a set of compact SSG international symbols by combining non-crystallographic point groups with the 230 crystallographic space groups.
The program first identifies all SSG operations and determines the SSG international symbol for a given magnetic system.
It then generates the SSG character tables of little groups at any $k$ point.
Finally, it computes the traces of matrix representations of SSG operations and assigns irreducible corepresentation labels to magnetic energy bands. The program is not only timely but also essential for advancing research on the study of magnons, altermagnetism, magnetic topology, and novel high-degeneracy excitations in SSG systems.

\ \textbf{Program summary} \\
{\it Program title: } IRSSG \\
{\it CPC Library link to program files:} \href{https://doi.org/10.17632/sjyxmpr9gw.1}{https://doi.org/10.17632/sjyxmpr9gw.1} \\
{\it Licensing provisions:} GNU General Public Licence 3.0\\
{\it Programming language:} Python 3, Fortran 90\\
{\it Nature of problem:} Determining the irreducible corepresentations of band structures under spin space groups (SSGs) is particularly valuable for band-character assignment, connectivity analysis, and topology diagnostics in SSG systems.\\
{\it Solution method:} The program identifies all SSG operations and determines the corresponding SSG international symbol. It constructs the $k$-little-group character tables using the Hamiltonian method, and determines the band irreducible corepresentations by comparing the traces of matrix representations of SSG operations acting on DFT wavefunctions or Wannier tight-binding states.
\end{abstract}

\maketitle
\vspace{-0.7\baselineskip}
\noindent\makebox[\textwidth][c]{\parbox{0.79\textwidth}{DOI: \href{https://doi.org/10.1016/j.cpc.2026.110190}{10.1016/j.cpc.2026.110190}}}

\section{Introduction}
Symmetry plays a fundamental role in condensed matter and materials physics by facilitating the prediction of emergent phenomena and material properties, such as ferroelectricity~\cite{PhysRevLett.106.107204,zhang2023,PhysRevLett.134.016801}, phonons~\cite{galiffi2024,shu2024,PhysRevB.110.075414}, topological insulators~\cite{zhang_topological_2009,hsieh2012,elcoro2021,PhysRevB.111.L041117}, and superconductivity~\cite{PhysRevResearch.1.013012,PhysRevX.12.011021,doi:10.1126/sciadv.aaz8367}. Symmetry also provides powerful constraints that significantly simplify otherwise complex many-body problems. For crystals, the symmetries are categorized into 230 crystallographic space groups (CSGs). In magnetic materials, a variety of spin configurations (\eg, ferromagnetic, antiferromagnetic, and helical orders) can emerge. To describe such materials in the presence of spin–orbit coupling (SOC), the time-reversal operation $\mathcal{T}$ is combined with the 230 CSGs, resulting in 1651 Shubnikov magnetic space groups (MSGs)~\cite{RevModPhys.40.359}. However, in the weak-SOC regime relevant for many magnetic materials and magnons, the spin and lattice degrees of freedom largely decouple, leading to a richer symmetry structure beyond MSGs. Therefore, Brinkman and Elliott introduced the concept of spin space groups (SSGs), where spin and lattice operations are treated individually, without being rigidly tied~\cite{doi:10.1098/rspa.1966.0211}. Later, Litvin provided a mathematically rigorous formulation of SSGs~\cite{LITVIN1974538} and systematically enumerated their corresponding spin point groups~\cite{litvin1977spin}. In this SOC-free setting, Liu \textit{et al.} provided a modern formulation of spin-group symmetry and clarified its implications for condensed-matter Hamiltonians and applications~\cite{PhysRevX.12.021016}. Building on these developments, several recent works have achieved a comprehensive enumeration and classification of SSGs~\cite{PhysRevX.14.031037,PhysRevX.14.031038,PhysRevX.14.031039}.

The SSGs are now widely used in the analysis of topological electronic bands, magnon spectra~\cite{PhysRevB.105.064430,chen2025}, superconducting order parameters~\cite{PhysRevB.111.054520}, altermagnetism~\cite{PhysRevX.12.031042,PhysRevX.12.040501}, anomalous Hall responses~\cite{PhysRevX.15.031006}, {\it etc.} 
Jiang \ea\ systematically enumerated all such SSGs, including 1421 collinear groups, 24,788 coplanar noncollinear groups, and 157,289 noncoplanar groups, in which the magnetic primitive cell is at most twelve times the atomic primitive cell~\cite{PhysRevX.14.031039}.  In addition, Chen \ea \ proposed an international symbol for SSGs. However, these Chen-Liu symbols can be rather involved because they are constructed within a classification scheme that divides SSGs into three types~\cite{PhysRevX.14.031038}. Later, they further developed an oriented international notation for SSGs, which explicitly encodes the absolute orientation of magnetic moments~\cite{liu2026symmetryclassificationmagneticorders}. Nonetheless, since this notation is also based on the same three-type classification, its explicit form may vary substantially from case to case, leading to additional complexity.

For nonmagnetic and magnetic topological materials, the program {\ttfamily IRVSP/IR2TB}~\cite{GAO2021107760}  represents the first computational tool capable of determining irreducible (co)representations [(co)irreps] and uncovering the band topology of electronic band structures computed using density functional theory (DFT) codes such as VASP~\cite{PhysRevB.54.11169,KRESSE199615}, Quantum ESPRESSO~\cite{Giannozzi_2009,Giannozzi_2017}, as well as any other code that has an interface to Wannier90~\cite{RevModPhys.84.1419,MOSTOFI20142309}. 
It can generate the input files needed to calculate elementary band representations and symmetry-based indicators via the {\ttfamily TopMat}~\cite{PhysRevB.106.035150} or {\ttfamily Check\_Topological\_Mat} server~\cite{vergniory_complete_2019}. Within this band-representation framework, methods have also been developed to diagnose unconventional materials with obstructed atomic limit via the {\ttfamily BRdecomp} server~\cite{GAO2022598}.
In this work, we develop  {\ttfamily IRSSG} to obtain the SSG coirreps of the band structures of magnetic materials with arbitrary SSG symmetries for the first time. {\ttfamily IRSSG} identifies all SSG operations and determines the SSG number and international symbol. Meanwhile, we introduce a set of compact SSG international symbols by combining non-crystallographic point groups with the 230 crystallographic space groups. 

The paper is structured as follows. In Section~\ref{concept}, we present the methods implemented in our program, including the basic concepts of SSGs as well as the method to construct coirreps based on SSG operations and to obtain matrix representations of materials. In Section~\ref{sec:implementation}, the general workflow of the {\ttfamily IRSSG} code is described in detail. The capabilities of {\ttfamily IRSSG} are described in Section~\ref{Capabilities of programs}. In Section~\ref{Installation and usage}, we introduce the installation steps and usage of the program. Some typical examples are shown in Section~\ref{Examples}. In Appendix~\ref{appendix:ssg_notation_compare}, we provide a side-by-side comparison between the symbols output by {\ttfamily IRSSG} and Chen-Liu symbols.

\section{Basic concepts of spin space groups and coirreps}
\label{concept}
\subsection{Spin space groups and international symbols}

Given a magnetic crystal structure with atoms $\{\boldsymbol{s}_i,\boldsymbol{r}_i,X_i\}$, SSG operations are defined as
\begin{equation}
  \begin{aligned}
    &\mathcal{G}=\{ \mathcal{O}_\alpha\equiv \{U_\alpha||R_\alpha|\bv_\alpha\}|U_\alpha\boldsymbol{s}_i=\boldsymbol{s}_j,R_\alpha\boldsymbol{r}_i+\bv_\alpha=\boldsymbol{r}_j, X_i=X_j \}, \\
    &U_\alpha\in O(3), 
    \quad
    \det(U_\alpha)=\begin{cases}
+1, & \mathcal{O}_\alpha\text{ is unitary without } \mathcal{T},\\
-1, & \mathcal{O}_\alpha\text{ is anti-unitary with } \mathcal{T},
\end{cases}
\\
& R_\alpha\in O(3), 
    \quad \det(R_\alpha)=\begin{cases}
+1, & R_\alpha\text{ is a proper rotation without } \mathcal{I},\\
-1, & R_\alpha\text{ is an improper rotation with } \mathcal{I}. 
\end{cases}
    \label{symm}
  \end{aligned}
\end{equation}
Here $\boldsymbol{s}_i$, $\boldsymbol{r}_i$, and $X_i$ are the magnetic moment, position, and element type of the $i$th atom, respectively. 
An SSG operation is denoted as ${\cal O}_\alpha=\{U_\alpha||R_\alpha|\bv_\alpha\}$, consisting of a spin part $U_\alpha$ and a lattice part $\{R_\alpha|\bv_\alpha\}$. 
The time-reversal ($\mathcal{T}$) symmetry reverses the spin direction and acts as an `inversion' symmetry in spin space. Therefore, $\det(U_\alpha)=-1$ indicates the presence of $\mathcal{T}$, being an anti-unitary symmetry. The $\det(R_\alpha)=-1$ indicates the presence of inversion ($\mathcal{I}$), resulting in an improper rotation. In an SSG, the spin rotation $U_\alpha$ and lattice rotation $R_\alpha$ are $O(3)$ matrices and independent. 
In a sense, the 230 CSGs and 1651 MSGs are special cases of SSGs.
Explicitly, a CSG is an SSG with $U_\alpha=E$ for all $\mathcal{O}_\alpha \in\mathcal{G}$, while an MSG is an SSG with $U_\alpha=\pm R_\alpha$ (sharing the same proper rotation) for all $\mathcal{O}_\alpha \in\mathcal{G}$. 
Therefore, although {\ttfamily IRSSG} is developed for SSGs, it also works for the 230 CSGs and 1651 MSGs.

In general, a special subgroup of $\mathcal{G}$ is defined as the spin-only group,
\begin{equation}
    \begin{aligned}
        \mathcal{S}_0&\equiv\{\mathcal{O}_\alpha\in\mathcal{G}|\{R_\alpha|\bv_\alpha\}=\{E|\boldsymbol{0}\}\},~~~~~ \mathcal{G}_0=\mathcal{G}/\mathcal{S}_0,\\
        \mathcal{G}&=\mathcal{S}_0\times \mathcal{G}_0 
    \end{aligned}
    \label{outS0}
\end{equation}
which contains the SSG operations whose lattice part is the identity. Since the coset representatives of $\mathcal{G}/\mathcal{S}_0$ form a group ($\mathcal{G}_0$), $\mathcal{G}$ can be expressed as a direct product of $\mathcal{S}_0$ and $\mathcal{G}_0$. 
According to the magnetic configuration, SSGs can be classified into three types: collinear (I), coplanar (II) and noncoplanar (III), each corresponding to a specific spin-only group, as shown in Table~\ref{SSG structure}. 
We now introduce two groups derived from $\mathcal{G}_0$: the spin part group $\mathcal{P}$ and the lattice part group $\mathcal{H}$,
\begin{equation}
    \mathcal{P}\equiv \{U_\alpha|~\forall \mathcal{O}_\alpha\in\mathcal{G}_0\},~~~~
    \mathcal{H}\equiv \{\{R_\alpha|\bv_\alpha\}|~\forall\mathcal{O}_\alpha\in\mathcal{G}_0\}.
    \label{PH}
\end{equation}
Clearly, $\mathcal{P}$ belongs to a non-crystallographic point group (NPG), while $\mathcal{H}$ belongs to a CSG. Thus, any SSG operation of $\mathcal{G}_0$ can be expressed by an element (x) of $\mathcal{H}$, with a superscript element ($\delta$) of $\mathcal{P}$ (\ie x$^\delta\equiv \{\delta||\rm x\}$).

Based on the above analysis of the SSG group structure, we propose a set of compact and intuitive SSG international symbols by the following steps: (1) identify the type of magnetic configuration (I: collinear; II: coplanar; III: noncoplanar) and the specific spin directions;
(2) identify the spin part group $\mathcal{P}$ (third column of Table~\ref{SSG structure}); (3) identify the lattice part group $\mathcal{H}$ with the international symbol Xxyz. The typical translations of the Bravais-lattice type X and the rotations x, y, z are the generators of $\mathcal{H}$; (4) find $\alpha,\beta,\gamma$ elements of $\mathcal{P}$ combined with three typical translational generators; (5) find $\delta/\epsilon/\eta$ element in $\mathcal{P}$ combined with the rotational generators x/y/z of $\mathcal{H}$. Thus, an SSG international symbol is given by $\rm X^{\alpha,\beta,\gamma}x^{\delta}y^{\epsilon}z^{\eta}(\mathcal{P}^{I/II/III})$. 
All SSG international symbols are listed on Huairou Symmetry Platform (HSP; \href{https://cmpdc.iphy.ac.cn/hsp/}{https://cmpdc.iphy.ac.cn/hsp/}). Since the spin space is decoupled from the lattice space, the axes in spin space have to be defined explicitly. Taking coplanar magnetic Mn$_3$Sn as an example, the SSG international symbol and the $x' y'$ axes of the coplanar spin plane are output by {\ttfamily IRSSG}.



\begin{table}[!t]
  \centering
  \caption{Spin space groups and international symbols. 
  The non-crystallographic point group $\mathcal{P}$ (in Schönflies notation) is the group formed by the spin parts of the elements of $\mathcal{G}_0$ in Eq.~(\ref{PH}). Its elements are listed as $\alpha$, $\beta$, $\gamma$, $\delta$, \ea, denoted in Hermann–Mauguin notation.
  The crystallographic space group $\mathcal{H}$ is the group formed by the lattice parts of $\mathcal{G}_0$ in Eq.~(\ref{PH}), whose international notation is $\rm Xxyz$. $\rm X$ stands for the Bravais-lattice type and $\rm x/y/z$ are the Hermann–Mauguin symbols of the rotational generators. A compact SSG international symbol can be expressed in the form of $\rm X^{\alpha,\beta,\gamma}x^{\delta}y^{\epsilon}z^{\eta}(\mathcal{P}^{I/II/III})$.}
  \begin{tabular}{|p{0.133\linewidth} |C{0.155\linewidth}|C{0.26\linewidth}|C{0.15\linewidth}|C{0.09\linewidth}|C{0.08\linewidth}|C{0.067\linewidth}|}
  \hline\hline
    \multicolumn{7}{|c|}{Spin space group $\mathcal{G}$ and international symbol $\rm X^{\alpha,\beta,\gamma}x^{\delta}y^{\epsilon}z^{\eta}(\mathcal{P}^{I/II/III})$}\\
    \multicolumn{7}{|c|}{ ($\mathcal{G}=\mathcal{S}_0\times \mathcal{G}_0$)}\\
    \hline
    \multirow[t]{4}{=}{Magnetic Configuration} &
    \multirow[t]{4}{=}{\qquad \qquad \qquad Spin-only group $\mathcal{S}_0$} &
    \multicolumn{5}{c|}{Group $\mathcal{G}_0$}\\
    \cline{3-7}
    & & {$\mathcal{P}$ (spin part)} &
      \multicolumn{2}{c|}{$\mathcal{H}$ (lattice part)} &
      \multicolumn{2}{c|}{$\mathcal{G}_0$ generators}\\
      \cline{4-7}
    & & ($\mathcal{P}\equiv \{\alpha,~\beta,~\gamma,~\delta,\cdots\}$)
      & ${\rm X}:\boldsymbol{a}_1,\boldsymbol{a}_2,\boldsymbol{a}_3$ & $\rm x/y/z$ & $\rm X^{\alpha,\beta,\gamma}$ & $\rm x^{\delta}$\\
    \hline
    I:\quad Collinear  &
     $\{\{C_{\infty z}||E|\boldsymbol{0}\},\allowbreak\ 
      \{M_x C_{\infty z}||E|\boldsymbol{0}\}\}$ &
    $C_1,~C_i$&
    \multirow[t]{3}{=}{P: $100,010,001$\quad C: $\frac{1}{2}\frac{1}{2}0,\bar{\frac{1}{2}}\frac{1}{2}0,001$\quad I: $\bar{\frac{1}{2}}\frac{1}{2}\frac{1}{2},\frac{1}{2}\bar{\frac{1}{2}}\frac{1}{2},\frac{1}{2}\frac{1}{2}\bar{\frac{1}{2}}$\quad A: $100,0\frac{1}{2}\bar{\frac{1}{2}},0\bar{\frac{1}{2}}\frac{1}{2}$\quad F: $0\frac{1}{2}\frac{1}{2},\frac{1}{2}0\frac{1}{2},\frac{1}{2}\frac{1}{2}0$\quad R: $\frac{2}{3}\bar{\frac{1}{3}}\bar{\frac{1}{3}},\frac{1}{3}\frac{1}{3}\bar{\frac{2}{3}},\frac{1}{3}\frac{1}{3}\frac{1}{3}$} &
    \multirow[t]{3}{=}{$1$, $\bar{1}$, $2$, $2_1$, $3$, $\bar{3}$, $3_1$, $3_2$, $4$, $\bar{4}$, $4_1$, $4_2$, $4_3$, $6$, $\bar{6}$, $6_1$, $6_2$, $6_3$, $6_4$, $6_5$, $a$, $b$, $c$, $d$, $e$, $m$, $n$}&
    \multirow[t]{3}{=}{$\{\alpha||E|\boldsymbol{a}_1\}$, $\{\beta||E|\boldsymbol{a}_2\}$, $\{\gamma||E|\boldsymbol{a}_3\}$} &
    $\{\delta||\rm x\}$\\
    \cline{1-3}
    II:~~Coplanar  & 
     $\{\{E||E|\boldsymbol{0}\}, \allowbreak
      \{M_z||E|\boldsymbol{0}\}\}$& $C_1,~C_i,~C_s,~C_n,~C_{nv}$ \qquad ($n\geq2$)&& & & \\
      \cline{1-3}
    III:~Noncoplanar  &  $\{\{E||E|\boldsymbol{0}\}\}$& $C_1,~C_i,~C_s,~C_n,~C_{nv},~C_{nh}$, $S_{2n}$, $D_{n}$, $D_{nh}$, $D_{nd}$, $T$, $T_h$, $T_d$, $O$, $O_h$, $I$, $I_h$& & & & \\
    \hline\hline
  \end{tabular}
  \label{SSG structure}
\end{table}

\subsection{Coirreps of spin space groups}
Since the magnetic translation group $\mathcal{T}_0 \equiv \{\mathcal{O}_\alpha \in\mathcal{G}_0 |U_\alpha=E,R_{\alpha}=E \}=\{\{E||E|\lambda\bt_1+\mu\bt_2+\nu\bt_3\}|~\lambda,\mu,\nu\in\mathbb{Z}\}$ is a normal subgroup of $\mathcal{G}$ ($\bt_{i=1,2,3}$ are the magnetic primitive lattice vectors), Bloch's theorem applies and the crystal wave vector $\bk$ is a good quantum number. Thus, coirreps of an SSG are classified by $k$ vectors. One can use a coirrep of the $k$-little group to construct the coirrep of the entire SSG, together with the $k$-star states. 
The little group of $\bk$, denoted as $LG(\bk)$, is defined as
\begin{equation}
LG(\bk) = \left\{ {\cal O}_\alpha \in\mathcal{G}\;\middle|\; \det(U_\alpha)R_\alpha \bk = \bk + \bG \right\}, 
~\bG = l\bg_1+m\bg_2+n\bg_3,\; l,m,n \in \mathbb{Z},
\label{eq:lgk}
\end{equation}
where $\bg_{i=1,2,3}$ are the magnetic primitive reciprocal lattice vectors (\ie $\bg_i\cdot \bt_j=\delta_{ij}$). Hereafter, without specific assignment, the SSG is typically referred to as the little group $LG(\mathbf{k})$.

To construct the linear coirreps of an SSG $\mathcal{G}$, we need to construct the projective coirreps of the quotient group $\mathcal{F}=\mathcal{G}/\mathcal{W}$, where $\mathcal{W}\equiv \mathcal{T}_0\times \{E, \bar E \}$ and $\bar E$ is a $2\pi$ rotation in spin space. The group $\mathcal{G}$ is an extension of $\mathcal{F}$ by $\mathcal{W}$. The quotient group can be expressed in terms of cosets as
\begin{equation}
\begin{aligned}
\mathcal{G}/\mathcal{W}\simeq \{g_i\mathcal{W}/\sim : g_i\in \mathcal{G}, g_i\sim g_ih,h\in \mathcal{W}\}, 
\end{aligned}
\end{equation}
where the elements of the quotient group (\ie $g_i$, $g_j$, and $g_ig_j$ are the coset representatives) satisfy
\begin{equation}
(g_i \mathcal{W}) (g_j \mathcal{W})= (g_ig_j \mathcal{W}),
\quad (g_i\mathcal{W})^{-1}=(g_{i}^{-1}\mathcal{W}),\quad e=(\mathcal{W}). 
\end{equation}
This induces a nontrivial factor system (Appendix~\ref{appendix:proof_factor}) for the projective representation $M$ of $\mathcal{F}$ ($g_i=\mathcal{O}_\alpha, ~g_j=\mathcal{O}_\beta $): 
\begin{equation}
\begin{aligned}
&M(\mathcal{O}_\alpha)\,M(\mathcal{O}_\beta)= \omega(\mathcal{O}_\alpha,\mathcal{O}_\beta)\,M(\mathcal{O}_\alpha \mathcal{O}_\beta ), \quad\omega({\cal O}_\alpha,{\cal O}_\beta)
= \omega_\tau({\cal O}_\alpha,{\cal O}_\beta)\,\omega_{s}({\cal O}_\alpha,{\cal O}_\beta).\\
&\omega_\tau({\cal O}_\alpha,{\cal O}_\beta) = \exp\!\left[-i \bk\cdot (R_\alpha-\det(U_\alpha))\bv_\beta\right],\quad \omega_{s}({\cal O}_\alpha,{\cal O}_\beta) = \pm 1.
\label{omega}
\end{aligned}
\end{equation}
The factor $\omega_{\tau}$ originates from the nontrivial phase induced by the translation subgroup $\mathcal{T}_0$ acting on Bloch states. The factor $\omega_s$ originates from the 2$\pi$ spin rotation group.
Once we obtain the complete factor system for $\mathcal{F}$, the projective coirreps $M({\cal O}_\alpha)$ of the quotient group $\mathcal{F}$ can be systematically constructed (Section~\ref{Construct irreducible corepresentations}), from which the double-valued linear representations of the full SSG $\mathcal{G}$ can be obtained~\cite{prbSSGReps}.

Since the SSG $\bk$-little group usually contains anti-unitary operations, we can decompose it as $LG(\bk)= \mathcal {L}+A\mathcal {L}$, where $\mathcal {L}$ is the unitary subgroup and $A$ is an anti-unitary element. The projective corepresentation (corep) is constructed from an irrep ($\Delta_a$) of the unitary subgroup $\mathcal{L}$ and its conjugate irrep ($\Delta^*_a$). Whether a projective corep is irreducible is determined by a criterion that is independent of the factor gauge~\cite{Yang_2021,10.1063/1.526419}:
\begin{equation} 
\frac{1}{|\mathcal{L}|}\sum_{\mathcal{O}\in\mathcal{L}}\frac{1}{2}\left\{\left|\text{Tr}[M(\mathcal{O})]\right|^2+\text{Tr}[M(A\mathcal{O})M^{*}(A\mathcal{O})]\right\}=1.
    \label{criterion}
\end{equation}
If the projective corep satisfies the condition in Eq.~(\ref{criterion}), then it is irreducible, being a projective coirrep. These projective coirreps can be classified into three cases by the torsion defined as
\begin{equation}
    R=\frac{1}{|\mathcal{L}|}\sum_{\mathcal{O}\in\mathcal{L}}|\chi(\mathcal{O})|^2=
\begin{cases}
1, & \text{real,\qquad \qquad case\ (a), labeled as }\Delta_a;\\
2, & \text{complex,\qquad case\ (b), labeled as }\Delta_a\Delta_b;\\
4, & \text{quaternionic,\ case\ (c), labeled as }\Delta_a\Delta_a.
\end{cases}
\end{equation}
In case (a), the projective coirrep is formed by a single projective irrep $\Delta_a$ of $\mathcal {L}$, being $\Delta_a$ coirrep.
In case (b), the projective coirrep is formed by the sum of two distinct projective irreps that constitute a complex-conjugate pair ($\Delta_b=\Delta_a^{*}$), being $\Delta_a\Delta_b$ coirrep.
In case (c), the projective coirrep is formed by the sum of two identical copies ($\Delta_a=\Delta_a^{*}$), being $\Delta_a\Delta_a$ coirrep.

\subsection{Traces of matrix representations of SSG operations}
A unitary operation acting on a spinor-valued function in real space is expressed as ${\cal O}f_{\sigma}(\br)= \sum_{\sigma'} Q_{\sigma\sigma'}f_{\sigma'}(\{R|\bv\}^{-1}\br)=\sum_{\sigma'} Q_{\sigma\sigma'}f_{\sigma'}( R^{-1}\br-R^{-1}\bv )$, where $Q$ is the SU(2) matrix of $U$.
The matrix representations, $O_{i}^{mn}$, can be obtained in the basis of the wavefunctions (WFs) $\ket{\psi_{n\bf k}}$ as $O_{\alpha}^{mn}=\braket{\psi_{m\bf k}}{{\cal O}_\alpha|\psi_{n\bf k}}$. The traces of the obtained matrix representations are essential for determining the corresponding irreps of $LG(\bk)$ and are defined as
\begin{equation}
 \text{Tr}[{\cal O}_\alpha]=\sum_{n}O_\alpha^{nn}\text{ with } O_\alpha^{nn}=\braket{\psi_{n\bf k}}{{\cal O}_\alpha|\psi_{n\bf k}},~ {\cal O}_\alpha\in LG(\bk).
\label{eq:mat}
\end{equation}

\subsubsection{Plane-wave basis}
In the plane-wave basis, the spinor WFs are expanded in plane waves as
\begin{align}
&\psi_{n\bf k}(\br)= \begin{pmatrix}\psi_{n\bf k\uparrow}(\br)\\ \psi_{n\bf k\downarrow}(\br)\end{pmatrix}=\sum_{j}\begin{pmatrix}C_{\uparrow j}^{n\bk}\\ C_{\downarrow j}^{n\bk}\end{pmatrix}  e^{i(\bk+\bG_j)\cdot \br}=\sum_{j,~\zeta= \{\up,\downarrow\} }C_{\zeta j}^{n\bk} e^{i(\bk+\bG_j)\cdot \br }\ket{\zeta }
\text{ with }\braket{\bk+\bG_i}{\bk+\bG_j}=\delta_{ij} 
\label{WFs}
\end{align}
The coefficients ($C_{\zeta j}^{n\bk}$) are obtained from \emph{ab initio} calculations and are output by DFT packages (\eg VASP and QE).
The action of SSG operations on the WFs is derived as
\begin{equation}
\begin{aligned}
{\cal O}_\alpha\psi_{n\bf k}(\br)&=\sum_{j}e^{i(\bk+\bG_j)\cdot (R^{-1}_\alpha\br-R^{-1}_\alpha\bv_\alpha)}Q_{\alpha}\begin{pmatrix}C_{\uparrow j}^{n\bk}\\ C_{\downarrow j}^{n\bk}\end{pmatrix}  \\
&=\sum_{j\zeta\zeta'}e^{iR_\alpha(\bk+\bG_j)\cdot (\br-\bv_\alpha)}Q_{\alpha,\zeta\zeta'} C_{\zeta' j}^{n\bk}\ket{\zeta}\\
&=\sum_{j\zeta\zeta'}e^{i(\bk+\bG_{j'})\cdot (\br-\bv_\alpha)}Q_{\alpha,\zeta\zeta'}C_{\zeta' j}^{n\bk}\ket{\zeta},\text{ with } \bk+\bG_{j'}\equiv R_\alpha(\bk+\bG_{j})\\
&=e^{-i\bk\cdot \bv_\alpha}\sum_{j\zeta\zeta'} e^{-i\bG_{j'}\cdot \bv_\alpha}e^{i(\bk+\bG_{j'})\cdot \br}Q_{\alpha,\zeta\zeta'}C_{\zeta' j}^{n\bk}\ket{\zeta}\text{ with } \bG_{j'}\equiv R_\alpha(\bk+\bG_{j})-\bk.
\end{aligned}
\end{equation}
Then, Eq.~(\ref{eq:mat}) can be written as
\begin{eqnarray}
\braket{\psi_{n\bf k}}{{\cal O}_\alpha|\psi_{n\bf k}}&=&e^{-i\bk\cdot \bv_\alpha} \sum_{j\zeta\zeta'} C_{\zeta j'}^{*n\bk} C_{\zeta' j}^{n\bk} e^{-i\bG_{j'}\cdot \bv_\alpha}Q_{\alpha,\zeta\zeta'},\text{ with } \bG_{j'}\equiv R_\alpha(\bk+\bG_{j})-\bk. 
\label{eq:pw}
\end{eqnarray}

\subsubsection{Localized Wannier basis}
In a tight-binding (TB) Hamiltonian, the spinor WFs are expanded in the basis of exponentially localized orthogonal orbitals: $\ket{\boldsymbol{0},\mu a \sigma}\equiv\phi_{a\sigma}^\mu(\br)\equiv \phi_{a}(\br-\tau_\mu)\ket{\sigma}$ and $\ket{\bL_j,\mu a\sigma}\equiv \phi_{a}(\br-\bL_j-\tau_\mu)\ket{\sigma}$, where $\mu$ labels the atoms, $a$ labels the orbitals, $\sigma$ labels the spins, $\bL_j$ labels the lattice vectors in 3D crystals, and $\tau_\mu$ labels the positions of atoms in the home unit cell.
At a given $k$ point, the WFs are given as
\begin{align}
&\psi_{n\bf k}(\br)= \begin{pmatrix}
    \psi_{n\vk\uparrow}(\br)\\\psi_{n\vk\downarrow}(\br)
\end{pmatrix}=\sum_{\mu a} \begin{pmatrix} C^{n\bk}_{\mu a\uparrow}\phi_{a \bk\uparrow}^\mu(\br)\\ C^{n\bk}_{\mu a\downarrow}\phi_{a \bk\downarrow}^\mu(\br)\end{pmatrix}
=\sum_{\mu a,~\sigma = \{\up,\downarrow\} }C^{n\bk}_{\mu a\sigma}\phi_{a \bk\sigma  }^\mu(\br)\text{, where $n$ is a band index},\notag  \\
&\phi_{a \bk\sigma}^\mu(\br)=\phi_{a \bk }^\mu(\br)\ket{\sigma }=\sum_j \phi_{a}(\br-\tau_\mu-\bL_j)e^{i \mathbf{k}\cdot (\bL_j+\tau_\mu)}\ket{\sigma  }, \braket{\phi_{b\bk\sigma'}^{\mu'}}{\phi_{a\bk\sigma}^{\mu}}=\delta_{\mu\mu'}\delta_{ab}\delta_{\sigma\sigma'}.\label{eq:fft}
\end{align}
The states $\phi_{a \bk\sigma}^\mu(\br)$ are the Fourier transforms of the local orbitals $\phi_{a\sigma}^\mu(\br)$, as shown in Eq.~(\ref{eq:fft}).
The coefficients are obtained as eigenvectors of the TB Hamiltonian
$H_{\mu'b\sigma',\mu a\sigma}(\bk)=\sum_j e^{i\bk\cdot (\bL_j+\tau_\mu-\tau_{\mu'})}\braket{{\bf{0}},\mu'b\sigma'}{\hat H|\bL_j, \mu a\sigma}$.
The action of rotational symmetries $R_\alpha$ on the local orbitals $\phi_{a\sigma}(\br)$ at the $\mu$ site is given by
\begin{align}
\widehat{R_\alpha\phi}_{a\sigma}(\br)\equiv R_\alpha\phi_{a\sigma}(\br)=\sum_b\phi_{b\sigma}(\br) D^{\alpha,\mu}_{ba}.
\label{eq:Drep}
\end{align}
These $D$-matrices are explicitly given in Table A.3 of Ref.~\cite{GAO2021107760}. In the basis of real spherical harmonic functions with different total angular momenta (integer $l$), these $D$-matrices are real.

The action of the SSG operation $\mathcal{O}_\alpha$ on the states $\psi_{n\mathbf{k}}(\br)$ is given by
\begin{align}
&{\cal O}_\alpha \psi_{n\mathbf{k}}(\br)\notag\\
=&\sum_{a\mu}Q_\alpha\begin{pmatrix}
    C_{\mu a\uparrow}^{n\vk}\phi_{a \mathbf{k}\uparrow}^{\mu}(R_\alpha^{-1}\br-R_\alpha^{-1}\bv_\alpha)\\C_{\mu a\downarrow}^{n\vk}\phi_{a \mathbf{k}\downarrow}^{\mu}(R_\alpha^{-1}\br-R_\alpha^{-1}\bv_\alpha)
\end{pmatrix} \notag \\
=&\sum_{a\mu\zeta\zeta'j}Q_{\alpha,\zeta\zeta'}C_{\mu a\zeta'}^{n\vk}\phi_{a \zeta'}(R_\alpha^{-1}\br-R_\alpha^{-1}\bv_\alpha-\tau_\mu-\bL_j)e^{i \mathbf{k}\cdot (\bL_j+\tau_\mu)}\ket{\zeta}  \notag \\
=& \sum_{a\mu\zeta\zeta'j}Q_{\alpha,\zeta\zeta'}C_{\mu a\zeta'}^{n\vk}\phi_{a \zeta'}(R_\alpha^{-1}\left[\br-\bv_\alpha-R_\alpha\tau_\mu-R_\alpha\bL_j\right])e^{i \mathbf{k}\cdot (\bL_j+\tau_\mu)}\ket{\zeta} \notag \\
=&\sum_{a\mu\zeta\zeta'j}Q_{\alpha,\zeta\zeta'}C_{\mu a\zeta'}^{n\vk}\widehat{R_\alpha\phi}_{a \zeta'}(\left[\br-\bv_\alpha-R_\alpha\tau_\mu-R_\alpha\bL_j\right])e^{i R_\alpha\mathbf{k}\cdot [R_\alpha(\bL_j+\tau_\mu)]}\ket{\zeta} \notag \\
=& e^{-i(R_\alpha\bk\cdot \bv_\alpha)}\sum_{a\mu\zeta\zeta'j}Q_{\alpha,\zeta\zeta'}C_{\mu a\zeta'}^{n\vk}\widehat{R_\alpha\phi}_{a \zeta'}(\left[\br-(\bv_\alpha+R_\alpha\tau_\mu)-R_\alpha\bL_j\right])e^{i R_\alpha\mathbf{k}\cdot [R_\alpha\bL_j+(R_\alpha\tau_\mu)+\bv_\alpha]}\ket{\zeta}\notag \\
=& e^{-i(R_\alpha\bk\cdot \bv_\alpha)}\sum_{a\mu\zeta\zeta'j}Q_{\alpha,\zeta\zeta'}C_{\mu a\zeta'}^{n\vk}\widehat{R_\alpha\phi}_{a \zeta'}(\br-(\tau_{\mu'}+\bL^i_0)-R_\alpha\bL_j)e^{i R_\alpha\mathbf{k}\cdot [R_\alpha\bL_j+(\tau_{\mu'}+\bL^i_0)]}\ket{\zeta} \text{ using }\bv_\alpha+R_\alpha\tau_\mu=\bL^i_0+\tau_{\mu'} \notag\\
=& e^{-i(R_\alpha\bk\cdot \bv_\alpha)}\sum_{a\mu\zeta\zeta'j}Q_{\alpha,\zeta\zeta'}C_{\mu a\zeta'}^{n\vk}\widehat{R_\alpha\phi}_{a \zeta'}(\br-\tau_{\mu'}-\bL_{j'})e^{i R_\alpha\mathbf{k}\cdot (\bL_{j'}+\tau_{\mu'})}\ket{\zeta}
\text{ with } \bL_{j'}=\bL^i_0 +R_\alpha\bL_j \notag \\
=& e^{-i(R_\alpha\bk\cdot \bv_\alpha)}\sum_{ab\mu\zeta\zeta'}Q_{\alpha,\zeta\zeta'}C_{\mu a\zeta'}^{n\vk} D^{\alpha,\mu}_{ba}\phi_{b,R_\alpha\mathbf{k}\zeta'}^{\mu'}(\br)\ket{\zeta}
\text{ with } \widehat{R_\alpha\phi}_{a\sigma}(\br)\equiv \sum_b\phi_{b\sigma}(\br) D^{\alpha,\mu}_{ba}.
\end{align}
Thus, Eq.~(\ref{eq:mat}) can be written as
\begin{eqnarray}
\braket{\psi_{n\bf k}}{{\cal O}_\alpha|\psi_{n\bf k}}&=&e^{-i(R_\alpha\bk\cdot \bv_\alpha)} \sum_{ab\mu\zeta\zeta'}  e^{i(R_\alpha\bk-\bk)\cdot \tau_{\mu'}} C^{n\bk*}_{\mu'b\zeta}  C^{n\vk}_{\mu a\zeta'} D^{\alpha,\mu}_{ba} Q_{\alpha,\zeta\zeta'},
 \text{ with } \bv_\alpha+R_\alpha\tau_\mu=\bL^i_0+\tau_{\mu'}. 
 \label{eq:tb}
\end{eqnarray}
In matrix form, 
\begin{eqnarray}
&\braket{\psi_{n\bf k}}{{\cal O}_\alpha|\psi_{n\bf k}}=e^{-i(R_\alpha\bk\cdot \bv_\alpha)}\left[\overline{C(\bk)^{\dagger}[V(R_\alpha\bk-\bk)\otimes \sigma_0](D^\alpha\otimes Q_\alpha)C(\bk)}\right]_{nn}, \notag 
 \\
&\text{ with } \overline{V}(\bk)_{\mu'b,\mu a}  = e^{i\bk\cdot {\tau_\mu}}\delta_{\mu\mu'}\delta_{ab},~\overline{C(\bk)}_{\mu a\zeta,n}=C^{n\bk}_{\mu a\zeta},~
\overline{D}^\alpha_{\mu'b,\mu a}=\begin{cases}
\begin{array}{cc}
D^{\alpha,\mu}_{ba}, &\text{if }\bv_\alpha+R_\alpha\tau_\mu=\bL^i_0+\tau_{\mu'};\\
                   0, &\text{otherwise}.
\end{array}\end{cases}
\label{eq:9}
\end{eqnarray}

Based on the above derivations, the code has been extended to work with the TB Hamiltonians. Thus, it works for any DFT code that has an interface to Wannier90, \eg VASP, Wien2k~\cite{wien2k,SCHWARZ200271} and OpenMX~\cite{PhysRevB.72.045121,PhysRevB.69.195113,PhysRevB.67.155108}. To run {\ttfamily IRSSG}, users must provide two input files: {\ttfamily case\_hr.dat} and {\ttfamily tbbox.in}. The file {\ttfamily case\_hr.dat} contains the TB parameters and can be generated by Wannier90 ~\cite{RevModPhys.84.1419,MOSTOFI20142309}, by users from a toy TB model, or from the Slater-Koster method~\cite{PhysRev.94.1498} or by discretization of a $k\cdot p$ model onto a lattice~\cite{willatzen2009kp}. The other input file, {\ttfamily tbbox.in}, should be consistent with the TB parameters. 
The {\ttfamily tbbox.in} file for Mn$_3$Sn with a type-II configuration is provided in Appendix~\ref{tbbox}.

\begin{figure}[!t]
  \centering
    \includegraphics[width=0.7\linewidth]{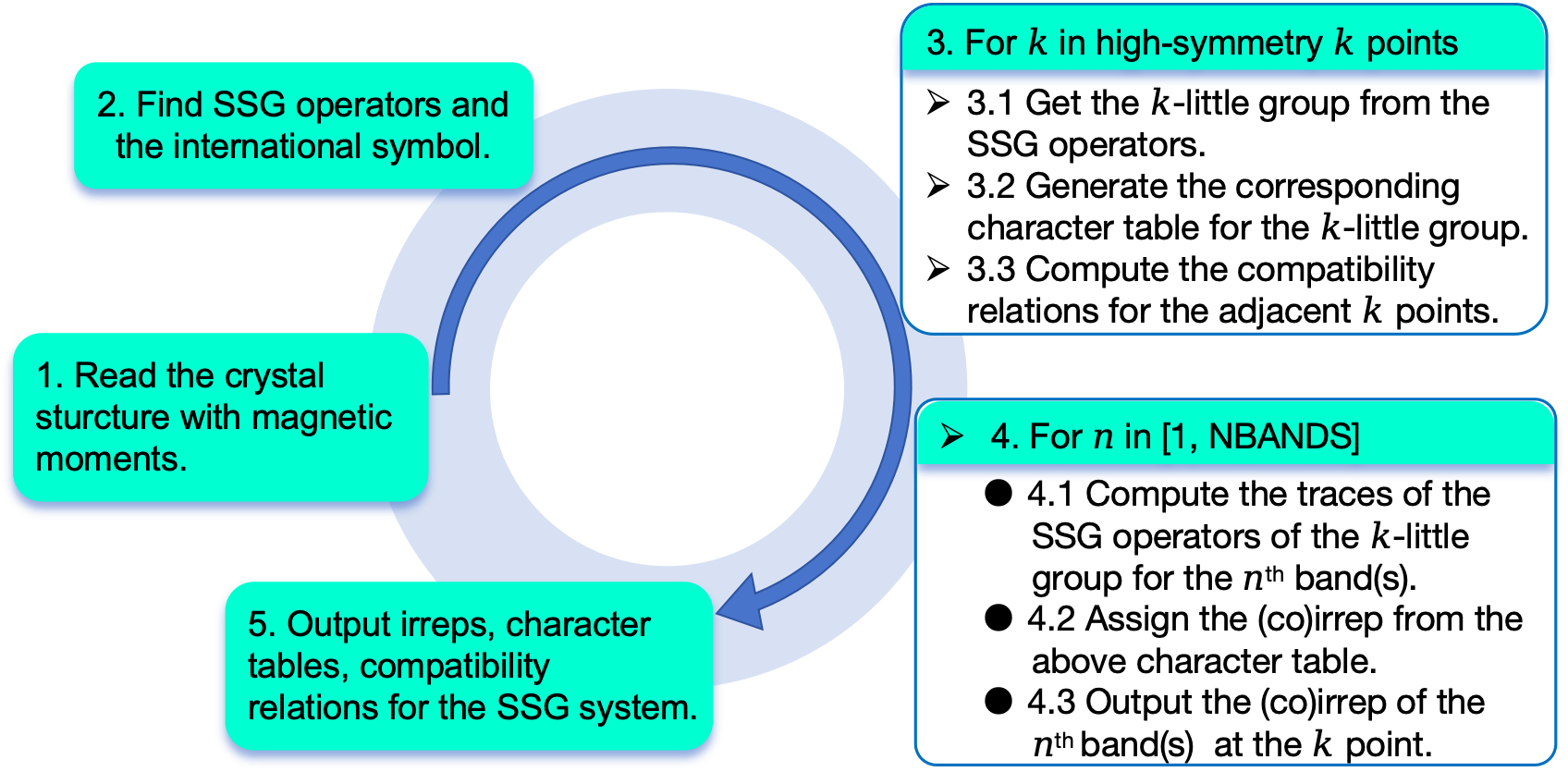}
  \caption{The workflow of {\ttfamily IRSSG} for spin space groups.}
  \label{fig:procedure}
\end{figure}
\begin{table}[!h]
  \centering
  \caption{Brief summary of key files.}
  \begin{tabular}{lll}
    \hline\hline
    File & Description & Input \\
    \hline
    \texttt{poscar\_io.py} & Reading \texttt{POSCAR} with magnetic moments. & 
    \texttt{\eg POSCAR} \\
    \texttt{find\_ssg\_operation.py} & Finding all operations in $\mathcal{G}_0$.& \\
    \texttt{get\_ssg\_number.py} & Obtaining the SSG number.&\\
    \texttt{get\_ssg\_symbol.py} & Obtaining the SSG international symbol.&\\
    \texttt{wave\_data.f90} & Reading the coefficients $C_{\zeta j}^{n\bk}$. & \texttt{\eg WAVECAR} \\
    \texttt{get\_ssg.f90}& Constructing all SSG operations in $\mathcal{G}$ [Eq.~(\ref{outS0})].&  \\
    \texttt{kgroup.f90} & Determining the $k$-little groups. & \\
    \texttt{linear\_rep.f90} & Constructing character tables at high-symmetry $k$ points& \\
    \texttt{chrct.f90} & Computing the traces [Eq.~(\ref{eq:pw})] and determining the coirreps. & \\
    \hline\hline
  \end{tabular}
  \label{keyfile}
\end{table}

\section{Implementations and features}\label{sec:implementation}
In this section, we focus on the implementation details of the {\ttfamily IRSSG} software package and introduce the major features of its component framework. {\ttfamily IRSSG} is developed to determine coirreps of magnetic energy bands for magnetic compounds from DFT calculations (taking VASP as an example in the main text).
The package first identifies all SSG operations and determines the SSG international symbol. 
It then generates the SSG character tables of the little groups at the 
$k$ points. Finally, it computes the traces of matrix representations of SSG operations and assigns the coirrep labels to magnetic energy bands. 
The main workflow is shown in Fig.~\ref{fig:procedure}, and a brief summary of the key files is given in Table~\ref{keyfile}.

\begin{table}[!b]
  \centering
  \caption{Structure of the tuple variable {\ttfamily cell} obtained by the function \texttt{read\_poscar} is given below ($N$ is the total number of atoms).}
  \begin{ruledtabular}
  \begin{tabular}{ccl}
    Index & Type & Description \\
    \hline
    0 & \texttt{numpy.ndarray[(3,3), float]} & lattice vectors ($\bt_1,\bt_2,\bt_3$) of the magnetic crystal structure \\
    1 & \texttt{numpy.ndarray[(N,3), float]} & fractional coordinates of atoms ($f_1\bt_1,f_2\bt_2,f_3\bt_3$) \\
    2 & \texttt{list[int]} & species indices of atoms \\
    3 & \texttt{list[str]} & element symbols of atoms \\
    4 & \texttt{numpy.ndarray[(N,3), float]} & magnetic moments of atoms in Cartesian coordinates
  \end{tabular}
  \end{ruledtabular}
  \label{tab:read_poscar_return}
\end{table}

\subsection{Reading the structure and magnetic moments}
\begin{figure}[!t]
  \centering
  \raisebox{0.5cm}{\begin{overpic}[width=0.5\linewidth]{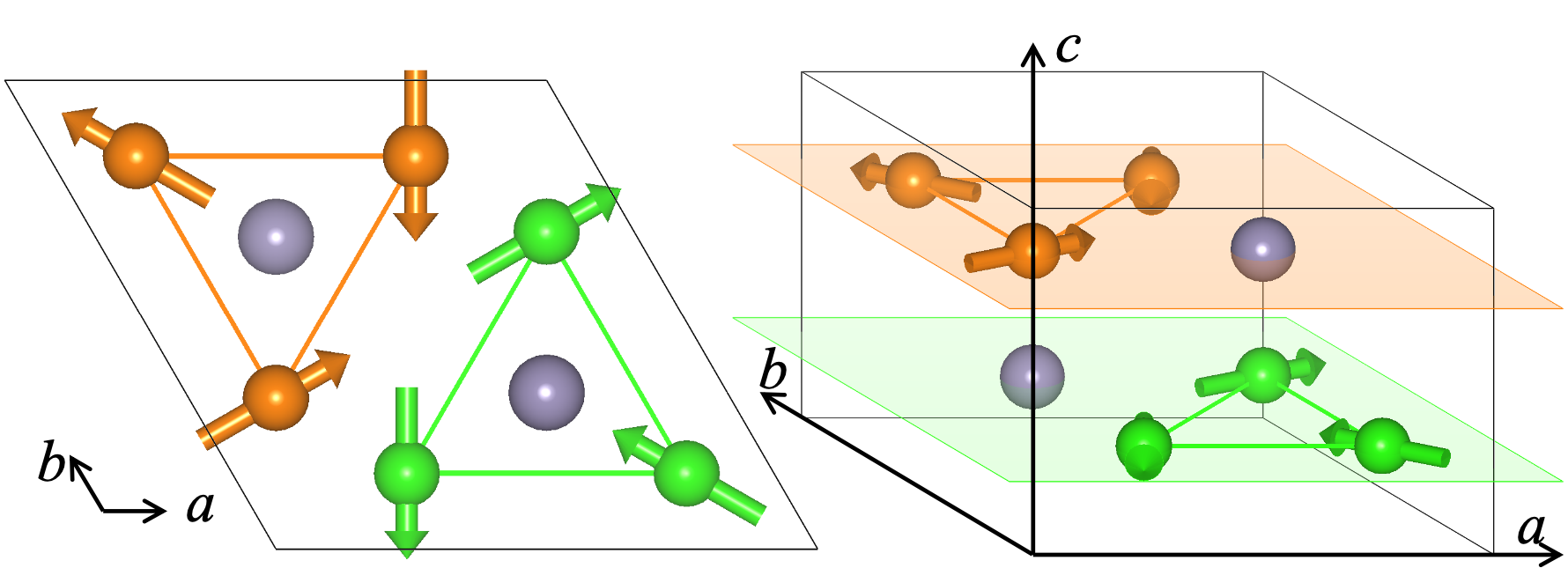}
    \put(-3,34){\small\textbf{(a)}}
  \end{overpic}}
  \hspace{0.3cm}
  \begin{overpic}[width=0.3\linewidth]{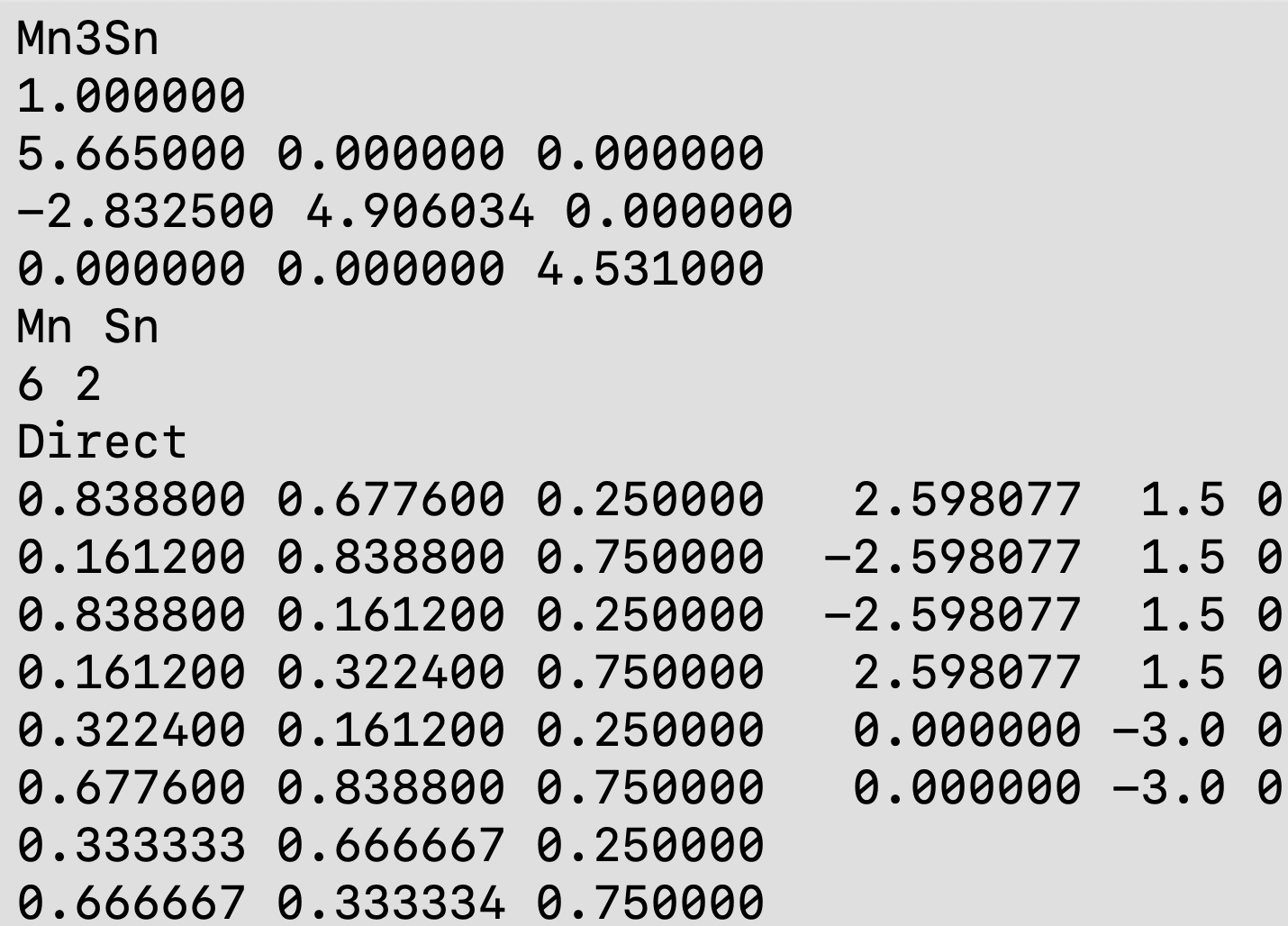}
    \put(-10,66){\small\textbf{(b)}}
  \end{overpic}
  \caption{(a) Magnetic crystal structure of Mn$_3$Sn in coplanar configuration.
  (b) The modified \texttt{POSCAR} including magnetic moments.}
  \label{structure}
\end{figure}

The function \texttt{read\_poscar} in \texttt{poscar\_io.py} reads the magnetic crystal structure, including lattice vectors, atomic positions, and magnetic moments.
The input is a modified {\ttfamily POSCAR} file, in which magnetic moments in Cartesian coordinates should follow the fractional coordinates of the atoms. A 5-tuple variable `{\ttfamily cell}' as defined in Table~\ref{tab:read_poscar_return} is obtained. 
For instance, the modified {\ttfamily POSCAR} file of Mn$_3$Sn with a type-II configuration is shown in Fig.~\ref{structure}(b).

\begin{figure}[!t]
 \includegraphics[width=0.6\linewidth]{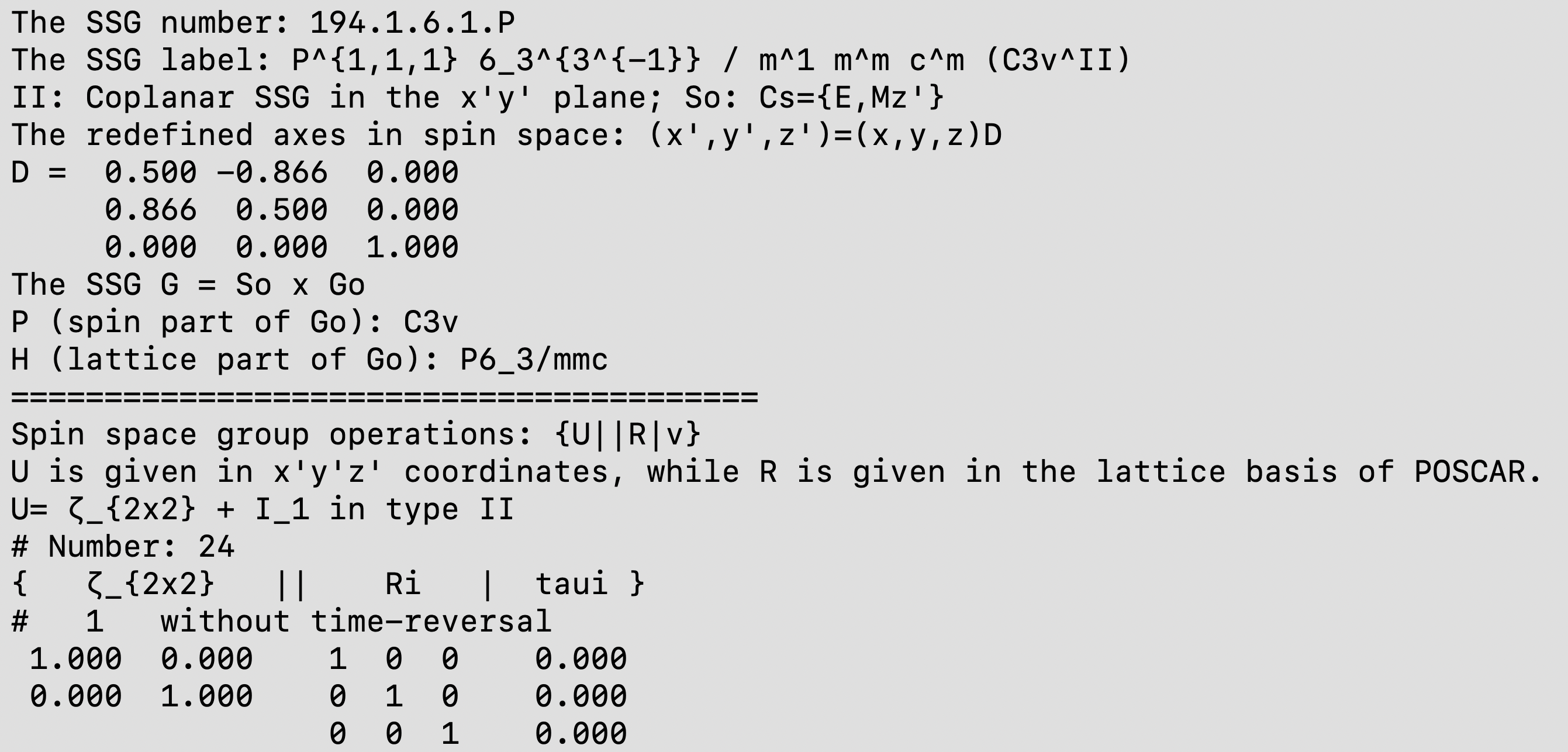}
 \caption{Screenshot of `ssg.out'. The SSG number and label are generated, and the redefined axes of spin space are also output explicitly.}
 \label{ssgout}
\end{figure}

\subsection{Finding SSG operations and international symbol}
The program {\ttfamily IRSSG} first determines the SSG magnetic configuration type (I: collinear; II: coplanar; III: noncoplanar). Then, it finds the SSG operations of the subgroup $\mathcal{G}_0$. Intuitively, all SSG elements of $\mathcal{G}_0$ must be contained in the direct product of the moment-part group $\mathcal{A}$ and the atomic-part group $\mathcal{B}$ (\ie $\mathcal{G}_0\leq\mathcal{A}\times\mathcal{B}$). The moment-part group $\mathcal{A}$ is an NPG determined solely by the magnetic moments. The atomic space group $\mathcal{B}$ is a CSG obtained by neglecting magnetic moments. All SSG operations are identified by {\ttfamily find\_ssg\_operation.py} using the {\ttfamily spglib}~\cite{Togo31122024} and {\ttfamily pymatgen}~\cite{ONG2013314} packages, and a dictionary variable {\ttfamily ssg\_ops} is generated.
Additionally, a binary file `ssg.data' is generated, which includes all operations of the full SSG $\mathcal{G}$. This file is required for constructing character tables and computing band representations.
Once `ssg.data' is prepared with the operations of a CSG or an MSG (as a specific SSG case), {\ttfamily IRSSG} works for them as well.

Then, based on all SSG operations, the SSG number and international symbol are determined by {\ttfamily get\_ssg\_number.py} and {\ttfamily get\_ssg\_symbol.py}, respectively.
Since the spin space and lattice space are independent, the redefined spin-space coordinates $(x',y',z')$ are output, depending on the spin configuration provided as input.
For type-I SSGs, the redefined $z'$ axis is chosen to align with the spin direction.
For type-II SSGs, the $x'y'$ plane is defined by the spin plane.
For type-III SSGs, the $x'y'z'$ axes are redefined following the standard NPG convention on the HSP.
Under the redefined coordinates, the spin space rotation becomes $U_\alpha = \zeta_\alpha \oplus I_1$ for type-I and $U_\alpha = I_2 \oplus \xi_\alpha$ for type-II. Here, $I_n$ is the $n$-dimensional identity, $\xi_\alpha=\pm 1$, and $\zeta_\alpha$ is a $2\times2$ $O(2)$ matrix.
A screenshot of `ssg.out' for Mn$_3$Sn with a coplanar configuration is shown in Fig.~\ref{ssgout}.

\subsection {Character tables at high-symmetry \texorpdfstring{$k$}{k} points}
We construct the character tables of the coirreps of the $k$-little groups $LG(\mathbf{k})$ at all $k$ points.

\begin{table*}[!b]
  \centering
  \caption{Brief summary of variables on SSG $\mathcal{G}$ and the little group $LG(\mathbf{k})$.}
  \begin{tabularx}{\textwidth}{p{1.3cm}p{5.2cm}p{1.5cm}X}
    \hline\hline
     & Variables & Type & Description \\
    \hline
    \multirow{4}{*}{SSG $\mathcal{G}$}
      & \texttt{num\_sym} & \texttt{integer} & Total number of operations in SSG $\mathcal{G}$ (\ie \ $\mathcal{G}/\mathcal{T}_0$). \\
      & \texttt{Spin\_rot(3,3,num\_sym)} & \texttt{real} & $U_{\alpha}$: spin part of operations $\mathcal{O}_{\alpha}\in\mathcal{G}$. \\
      & \texttt{Rot(3,3,num\_sym)} & \texttt{integer} & $R_{\alpha}$: rotation part of operations $\mathcal{O}_{\alpha}\in\mathcal{G}$. \\
      & \texttt{Tau(3,num\_sym)} & \texttt{real} & $v_{\alpha}$: translation part of operations $\mathcal{O}_{\alpha}\in\mathcal{G}$. \\
    \hline
    \multirow{5}{*}{$LG(\mathbf{k})$}
      & \texttt{NumLG} & \texttt{integer} & Total number of SSG operations in the little group $LG(\mathbf{k})$. \\
      & \texttt{litt\_group(:)} & \texttt{integer} & Indices of SSG operations belonging to $LG(\mathbf{k})$. \\
      & \texttt{aunt} & \texttt{integer} & Flag for (anti-)unitarity of $LG(\mathbf{k})$, 1: unitary, 2: anti-unitary. \\
      & \texttt{Nirrep} & \texttt{integer} & Number of irreps of the unitary group of $LG(\mathbf{k})$. \\
      & \texttt{Ch\_table1(Nirrep,NumLG/aunt)} & \texttt{complex} & Characters of the irreps for the unitary subgroup of $LG(\mathbf{k})$. \\
      \cline{2-4}
    
      & \texttt{Ncoirrep} & \texttt{integer} & Total number of coirreps of $LG(\mathbf{k})$. \\
      & \texttt{Ch\_table2(Ncoirrep,NumLG/aunt)} & \texttt{complex} & Characters of the unitary operations in the coirreps of $LG(\mathbf{k})$. \\
    \hline\hline
  \end{tabularx}
  \label{tab:ssg_vars_fourcol_fixed}
\end{table*}

\subsubsection{Obtaining the SSG \texorpdfstring{$k$}{k}-little group}
The SSG operations are read from `ssg.data' by {\ttfamily get\_ssg.f90}. 
An SSG operation that leaves $\bk$ invariant up to a reciprocal-lattice vector [Eq.~(\ref{eq:lgk})] belongs to the $k$-little group; this is implemented in {\ttfamily kgroup.f90}. All related variables are summarized in Table~\ref{tab:ssg_vars_fourcol_fixed} in detail.

\subsubsection{Generating the character table of the \texorpdfstring{$k$}{k}-little group}
\label{Construct irreducible corepresentations}
In this part, we use the Hamiltonian method to decompose the regular projective corep to obtain the projective coirreps, and finally obtain the linear coirreps, which are all done in file {\ttfamily linear\_rep.f90}. The related variable is listed in Table~\ref{tab:ssg_vars_fourcol_fixed}, while {\ttfamily Ncoirrep} and {\ttfamily Ch\_table2} are valid only when {\ttfamily aunt=2}.

The factor system $\omega$ is obtained from Eq.~(\ref{omega}). The regular projective corep is then constructed as
\begin{align}
M_{ij}(\mathcal{O}_\alpha)=
&\begin{cases}
\omega(\mathcal{O}_\alpha,\mathcal{O}_j)\delta_{\mathcal{O}_i,[\mathcal{O}_\alpha\mathcal{O}_j]}, & \mathcal{O}_\alpha\text{ is unitary},\\
\omega(\mathcal{O}_\alpha,\mathcal{O}_j)\delta_{\mathcal{O}_i,[\mathcal{O}_\alpha\mathcal{O}_j]}\mathcal{K}, & \mathcal{O}_\alpha\text{ is anti-unitary},
\end{cases}\notag\\
&\text{with }\mathcal{O}_\alpha,\mathcal{O}_i,\mathcal{O}_j,[\mathcal{O}_\alpha\mathcal{O}_j]\in LG(\bk)
\end{align}
where $\mathcal{K}$ is the complex conjugation. 

The regular projective corep is completely reducible and decomposes as the direct sum of all inequivalent projective coirreps, $M\cong \bigoplus_\rho d_\rho M^{(\rho)}$, where $M^{(\rho)}$ is the matrix of projective coirrep $\rho$, and $d_\rho$ is its multiplicity, equal to the dimension divided by the torsion of $\rho$. Thus, the projective coirreps are obtained by decomposing the regular projective corep using the Hamiltonian method. The results are written to `chart.dat'.

\subsubsection{Obtaining compatibility relations}
Due to subgroup relations between the SSG little groups of adjacent $k$ points, the compatibility relations are generated accordingly.
They are written to `chart.dat' as well.

\subsection {Computing the coirreps of magnetic energy bands} 

For each $k$ point, we obtain the coirreps of all magnetic energy bands.

\subsubsection{Computing the traces of the SSG operations in the \texorpdfstring{$k$}{k}-little group}

The coefficients $C_{\zeta j}^{n\bk}$ in Eq.~(\ref{WFs}) are stored in the \texttt{WAVECAR} generated by VASP and \texttt{wfc.dat} generated by QE. 
All coefficients are read by {\ttfamily wave\_data.f90} and stored in the complex variables {\ttfamily coeffa} ($C^{n\bk}_{\uparrow j}$) and {\ttfamily coeffb} ($C^{n\bk}_{\downarrow j}$).
To compute traces only, one can use the plane-wave wavefunctions directly.
The characters of SSG elements of $LG(\bk)$ are then computed using Eq.~(\ref{eq:pw}) for degenerate bands. In addition, {\ttfamily IRSSG} can compute the traces of SSG operations in the orthogonal localized Wannier basis via Eq.~(\ref{eq:tb}).

\subsubsection{Assigning coirreps to magnetic energy bands}
By comparing the obtained traces with the traces of the character tables, one can easily assign the coirreps, which is done in {\ttfamily chrct.f90}.

\subsection{Outputting coirreps, character tables, and compatibility relations for the SSG system}

\begin{figure}[!t]
  \centering
  \begin{overpic}[width=0.465\linewidth]{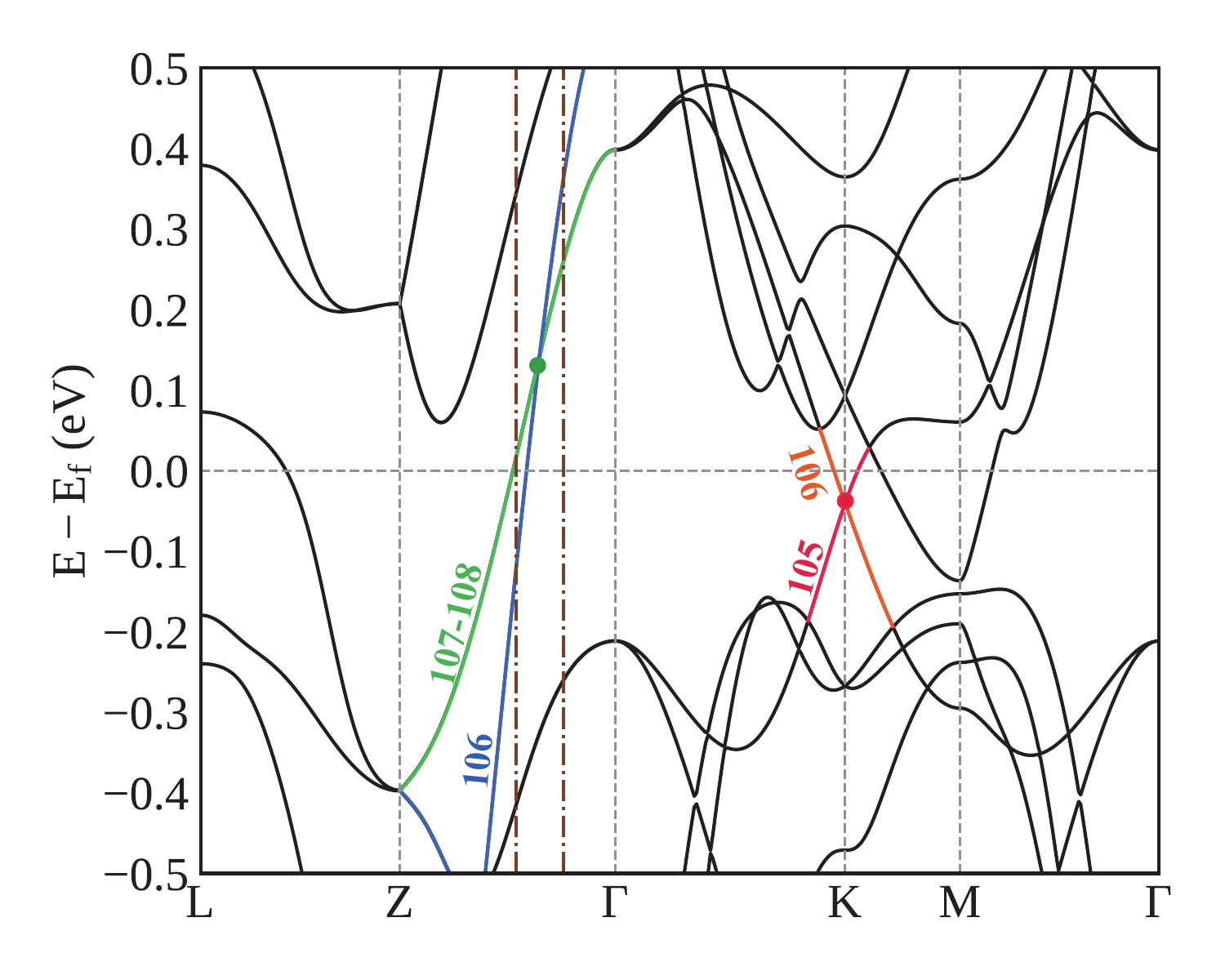}
    \put(3,73){\small\textbf{(a)}}
  \end{overpic}
  \hspace{0.24cm}
  \raisebox{0.4cm}{\begin{overpic}[width=0.49\linewidth]{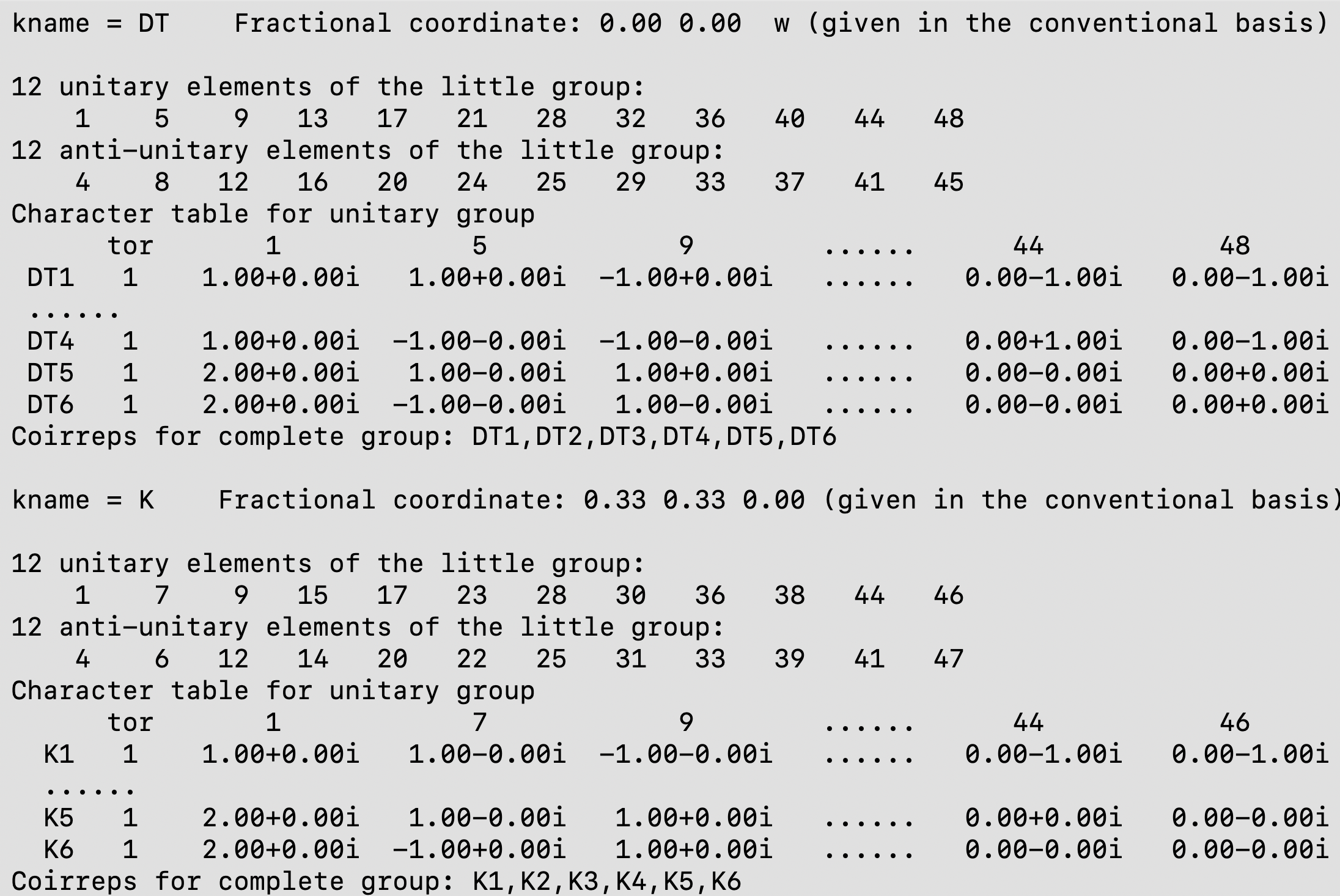}
    \put(-7,65){\small\textbf{(b)}}
  \end{overpic}}
  \caption{(a) Magnetic energy bands of Mn$_3$Sn with a type-II configuration.
  (b) Screenshot of `chart.dat'.}
  \label{Mn3Snbands}
\end{figure}

\begin{figure}[!t]
  \centering

  \begin{overpic}[width=0.7\linewidth]{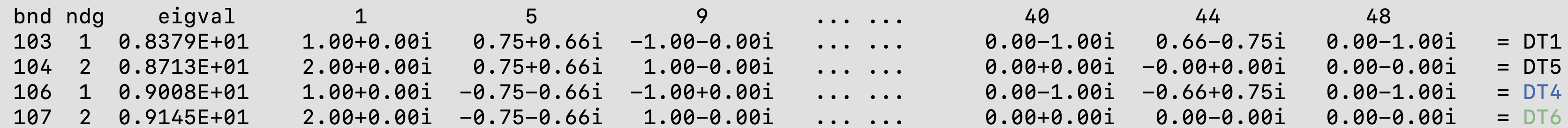}
    \put(-7,3){\small\textbf{(a)}}
  \end{overpic}

  \vspace{6pt} 

  \begin{overpic}[width=0.7\linewidth]{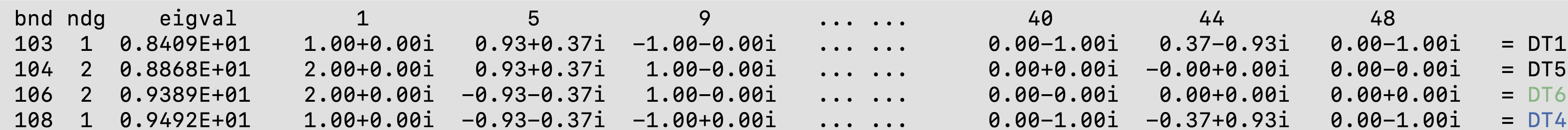}
    \put(-7,3){\small\textbf{(b)}}
  \end{overpic}

  \vspace{6pt}

  \begin{overpic}[width=0.7\linewidth]{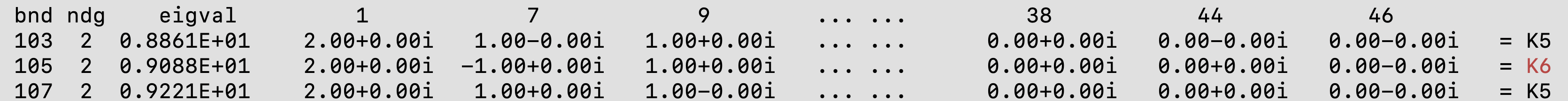}
    \put(-7,2){\small\textbf{(c)}}
  \end{overpic}

  \caption{The band coirreps determined by {\ttfamily IRSSG}, which are output in `irssg.out'. The first three columns give band indices, degeneracies, and energies (without subtracting the Fermi level $E_F$), respectively. Then the characters of unitary operations and coirrep labels are output. Panels (a)–(c) correspond to the two $\Delta$ (DT) points (two dash-dotted lines in Fig.~\ref{Mn3Snbands}(a)) and the K point.}
  \label{irssgout}
\end{figure}

The band characters and their coirreps are presented in the file `irssg.out'. Fig.~\ref{irssgout} shows excerpts of this file for Mn$_3$Sn with a type-II configuration. This file first lists all operations of $\mathcal{G}/\mathcal{T}_0$, each assigned an operation index. The character tables are output in the `chart.dat' file along with compatibility relations. This file presents the character tables for irreps of the unitary subgroups of $LG(\vk)$ at all $k$ points with distinct labels, and provides the compatibility relations between adjacent $k$ points. Fig.~\ref{Mn3Snbands}(b) shows a screenshot of the `chart.dat' file for Mn$_3$Sn with a type-II configuration.

\section{Capabilities of the program}
\label{Capabilities of programs}

Determining the representations of computed electronic bands is crucial for studying phenomena such as band crossings, high degeneracies, and $k\cdot p$ models and related effects. Tools such as {\ttfamily Phonopy}~\cite{phonopy-phono3py-JPCM,phonopy-phono3py-JPSJ} and {\ttfamily MOM2MSG}~\cite{MOM2MSG} can identify the CSG or MSG of materials. For a magnetic structure, {\ttfamily spinspg}~\cite{spinspg} and {\ttfamily findspingroup}~\cite{PhysRevX.14.031038} are available for identifying SSG operations and/or Chen-Liu symbol. However, tools for determining the (co)irreps of magnetic electronic states are lacking.
Although the complete character tables for CSGs and MSGs are available in the literature~\cite{AroyoPerezMatoCapillasKroumovaIvantchevMadariagaKirovWondratschek+2006+15+27,Aroyo:xo5013,aroyo2011crystallography}, the character tables of SSGs are not yet broadly accessible. Our program {\ttfamily IRSSG} can obtain the $k$-little group and generate character tables for SSGs. It also determines the coirreps of magnetic energy bands based on DFT calculations or Wannier functions. Meanwhile, with a proper `ssg.data', it also works for the 230 CSGs and 1651 MSGs. Thus, the corresponding `msg.data' is automatically generated for studying the magnetic system in the presence of SOC.
Furthermore, similar to {\ttfamily POS2MSG}~\cite{PhysRevB.106.035150}, which can generate the magnetic structures with highest MSG symmetries, we developed an auxiliary tool, {\ttfamily POS2SSG}, to generate highest-SSG-symmetry magnetic configurations by assigning magnetic moments to magnetic atoms, thereby producing structures with SSG numbers specified by the user.

\section{Installation and usage of {\ttfamily IRSSG}}
\label{Installation and usage}
{\ttfamily IRSSG} is an open-source package released on GitHub under the GNU General Public Licence 3.0, \href{https://www.gnu.org/licenses/gpl-3.0.html}{https://www.gnu.org/licenses/gpl-3.0.html}. It can be downloaded directly from the public code archive: \href{https://github.com/zjwang11/IRSSG}{https://github.com/zjwang11/IRSSG}. To build and install {\ttfamily IRSSG}, Python 3.8 or higher is required. The executable {\ttfamily irssg} and its dependencies are automatically downloaded and installed, and the coirreps of the $m$th--$n$th magnetic energy bands can be obtained by typing the following command:
\begin{cmdbox}
\$ pip install irssg\\
\$ irssg -ssg > ssg.out\\
\$ irssg -nb \$m \$n [ -tolE \$dE ] > irssg.out
\end{cmdbox}
\noindent where {\ttfamily -tolE} sets the energy tolerance $dE$ (0.001 eV by default) for identifying (near-)degenerate bands.
Here, we take coplanar Mn$_3$Sn as an example to illustrate the procedure. The magnetic configuration~\cite{P_J_Brown_1990}
is shown in Fig.~\ref{structure}(a). The corresponding modified \texttt{POSCAR} is shown in Fig.~\ref{structure}(b).
Then one can run the command `{\ttfamily irssg -ssg > ssg.out}', which generates the `ssg.out' (only the modified \texttt{POSCAR} is required as input), as shown in Fig.~\ref{ssgout}. From this output, one can find that the SSG international symbol for the Mn$_3$Sn configuration is $P^{1,1,1} 6_3^{3^{-1}} / m^1 m^m c^m (\mathrm{C_{3v}^{II}})$ with SSG number 194.1.6.1.P. In addition, the redefined coordinates in spin space, the spin part group $\mathcal{P}$ and the lattice part group $\mathcal{H}$ of the SSG are also presented. The elements ($\{U||R|\bv\}$) of the SSG are generated. 

We first perform the self-consistent and band-structure DFT calculations for the magnetic configurations (\eg setting `NONCOLLINEAR=.TRUE.; LSORBIT=.FALSE.' in \texttt{INCAR} for VASP). The wavefunctions must be output in the DFT calculation. The noncollinear band structure for Mn$_3$Sn is obtained in Fig.~\ref{Mn3Snbands}(a). We can find that there might be two `crossing' points around the Fermi level, which need to be confirmed by the SSG coirreps. Then one can use the command `{\ttfamily irssg -nb 103 108 > irssg.out}' to obtain the SSG coirreps of 103-108 -th bands, with the results written to `irssg.out'. The character tables (including compatibility relations) for all $k$ points are generated in `chart.dat', whose screenshot is given in Fig.~\ref{Mn3Snbands}(b). Explicitly, the screenshots of `irssg.out' are given in Fig.~\ref{irssgout} for the three $k$ points (denoted by two brown dash-dot lines in Fig.~\ref{Mn3Snbands}(a) and K), which indicate that the crossing at Z$\Gamma$ is a 3-fold crossing point, formed by DT4 (106-th band) and DT6 (107-th and 108-th bands). The Dirac point at K forms a K6 coirrep, indicating that it is a symmetry-enforced, conventional twofold band-degeneracy point~\cite{PhysRevX.12.021016}. The coirrep matrices of the generators of $LG({\mathrm{K}})$ are $D(\{E||M_z|0,0,\frac{1}{2}\})=-\sigma_0$, $D(\{C^{-1}_{3z}||C_{3z}|0,0,0\})=\frac{1}{2}(\sigma_0-\sqrt{3}i\sigma_z)$, $D(\{C_{2(\frac{\sqrt{3}}{2},\frac{1}{2},0)}||M_y|0,0,\frac{1}{2}\})=-i\sigma_x$, and $D(\{M_z||P|0,0,0\})=\sigma_x\mathcal{K}$. Based on the coirrep matrices of the generators, the $k\cdot p$ effective model is also obtained for the SSG system via the invariant theory
\begin{equation}
H(k_x,k_y,k_z)=\begin{pmatrix}
0 &
b_1 k_- + c_2 k_+^2 \\[6pt]
b_1 k_++c_2 k_-^2 &
0
\end{pmatrix} + [a_1+c_1(k_x^2+k_y^2)+c_3 k_z^2] \sigma_0
\end{equation}
where $k_\pm = k_x \pm i k_y$, and the $k\cdot p$ parameters are $a_1 = 9.0882\rm eV$, $b_1 = 1.0768\rm eV\cdot\AA$, $c_1 = 1.9892\rm eV\cdot\AA^2$, $c_2 = 1.3385\rm eV\cdot\AA^2$ and $c_3 = -11.9826\rm eV\cdot\AA^2$. The parameters are directly obtained by using {\ttfamily VASP2KP}~\cite{Zhang_2023} in the DFT calculation. One can clearly see that it is a quasi-2D-Dirac point with linear dispersions in the $xy$ plane, while it is quadratic (doubly degenerate) in the $z$ direction. Only two Pauli matrices are allowed in the $k\cdot p$ Hamiltonian. A nontrivial $\pi$ Berry phase of the circle surrounding it can be obtained by the 1D Wilson loop method.

\begin{figure}[!t]
  \centering
  \begin{overpic}[width=0.34\linewidth]{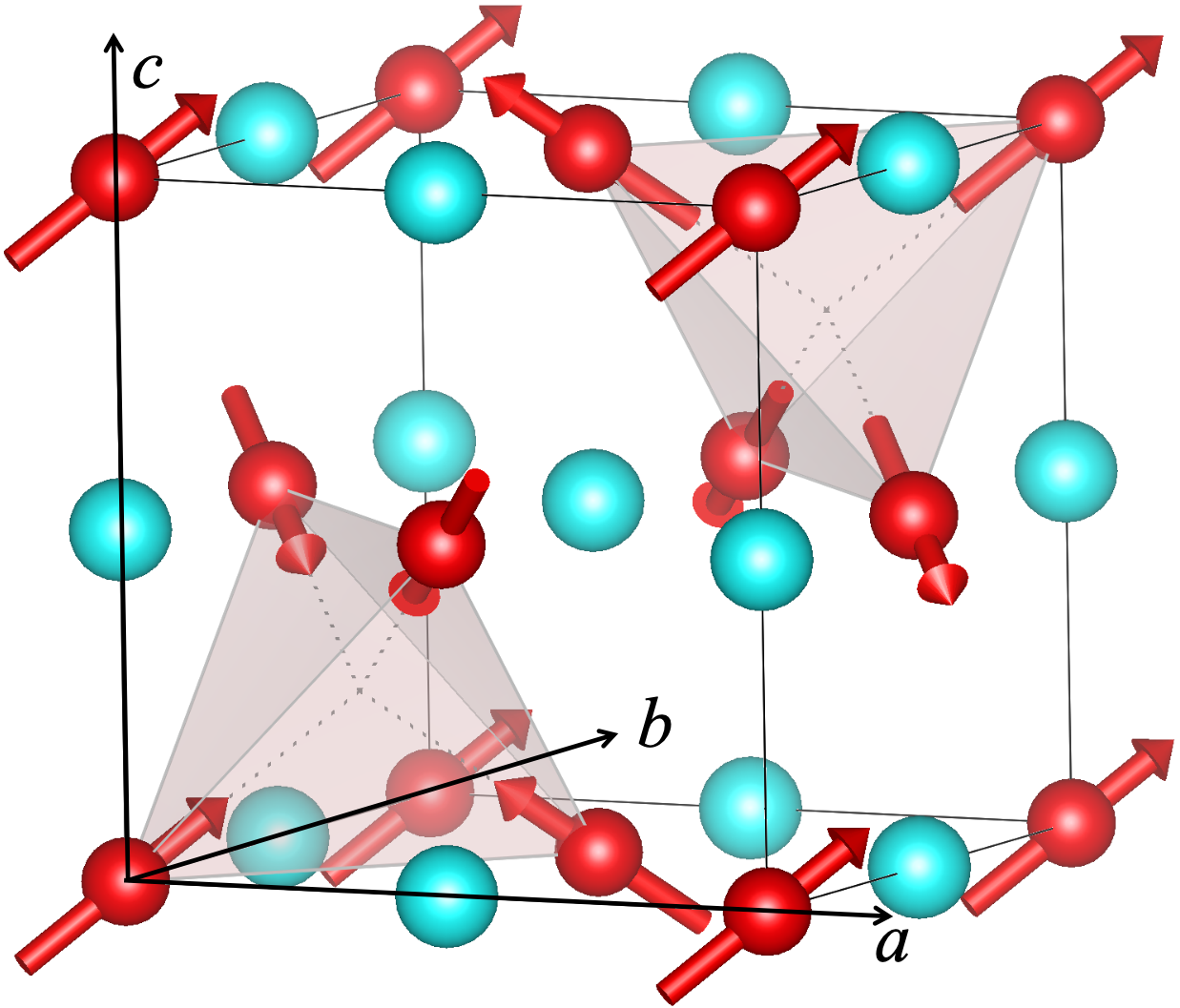}
    \put(0,75){\small\textbf{(a)}}
  \end{overpic}
  \begin{overpic}[width=0.35\linewidth]{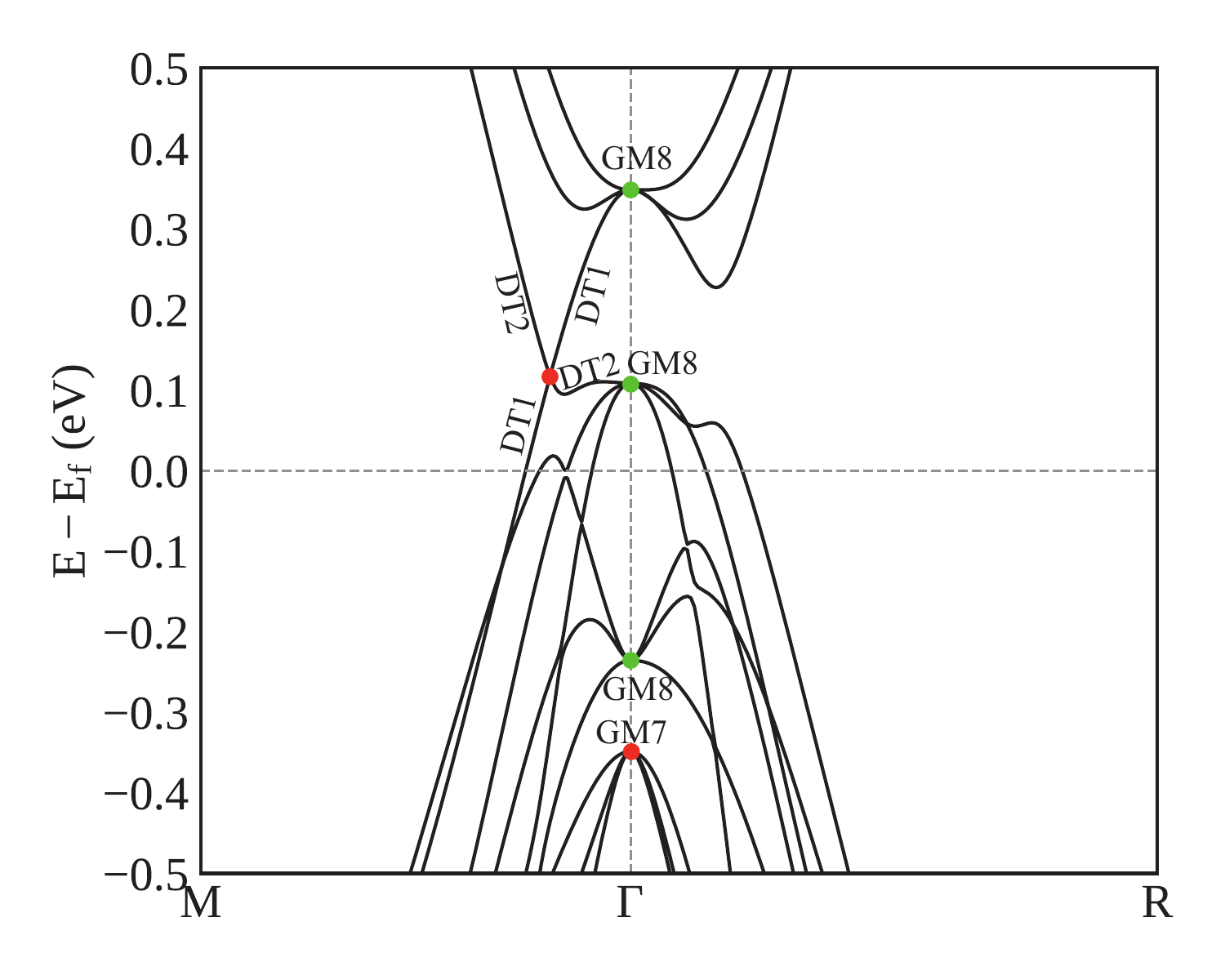}
    \put(0,73){\small\textbf{(b)}}
  \end{overpic}
  \begin{overpic}[width=0.29\linewidth]{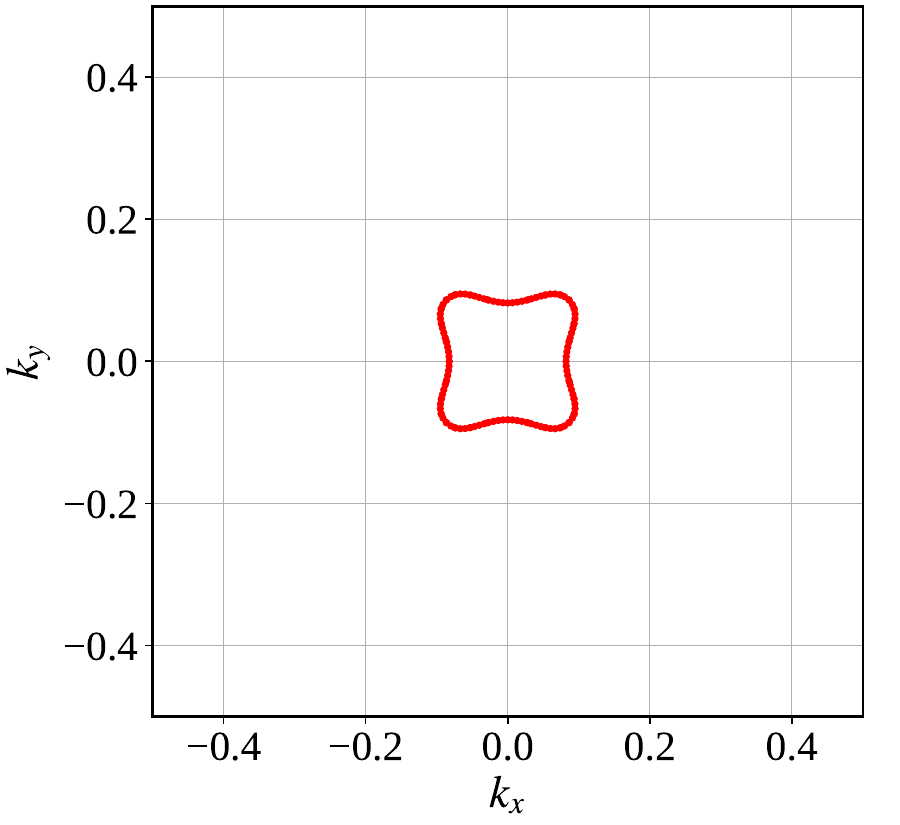}
    \put(0,88){\small\textbf{(c)}}
  \end{overpic}

  \par\medskip

\noindent\hspace*{-1.7cm}%
\begin{minipage}{\linewidth}
  \begin{overpic}[width=0.80\linewidth]{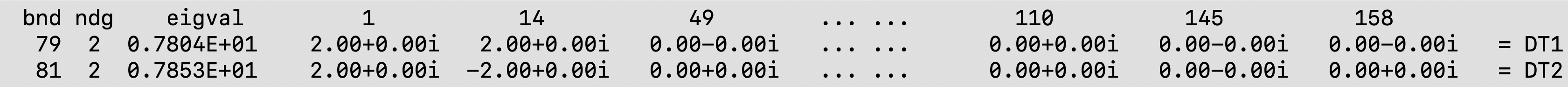}
    \put(-6,2){\small\textbf{(d)}}
  \end{overpic}
\end{minipage}
  \par\medskip
  
\noindent\hspace*{-8cm}%
\begin{minipage}{\linewidth}
  \begin{overpic}[width=0.45\linewidth]{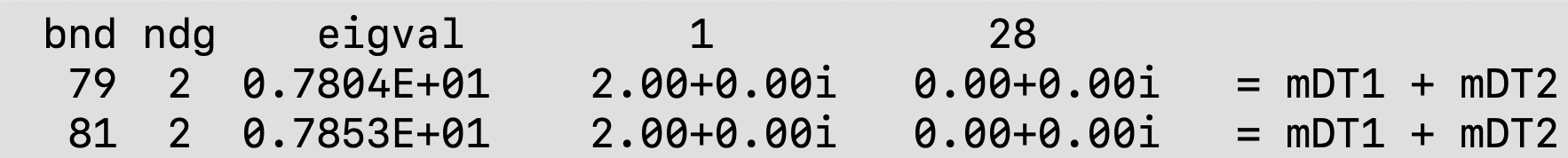}
    \put(-10,4){\small\textbf{(e)}}
  \end{overpic}
\end{minipage}

  \caption{(a) Magnetic crystal structure of NpBi with a type-III configuration, where Np atoms are magnetic.
  (b) Magnetic energy bands of NpBi.
  (c) The fourfold nodal line of NpBi in the $k_z=0$ plane. SSG coirreps (d) and MSG coirreps (e) of the related two doubly degenerate bands. Being different from SSG coirreps, the `m' is added as the prefix of MSG coirreps.}
  \label{NpBi}
\end{figure}
\section{Applications to magnetic materials}
\label{Examples}
\subsection{The fourfold nodal lines in NpBi}
The noncoplanar magnetic configuration of NpBi~\cite{BURLET1992131} is illustrated in Fig.~\ref{NpBi}(a). Its SSG international symbol $F^{2,2,2} m^1 \bar{3}^{3^{-1}} m^m ({\rm T_d^{III}})$ and SSG number 225.4.6.2 are obtained by {\ttfamily IRSSG}. After the noncollinear DFT calculations in the absence of SOC, 
the band structure is obtained along M$\Gamma$R (Fig.~\ref{NpBi}(b)) and the wavefunctions (\eg \texttt{WAVECAR} in VASP) are written out as well, with the total number of electrons being 80. Using the command `{\ttfamily irssg -nb 63 86 > irssg.out}', the coirreps are generated for 63-86 -th bands. We find that the crossing along M$\Gamma$ is formed by DT1 and DT2. The crossing is fourfold degenerate. In addition, the fourfold crossing is part of the fourfold nodal line in the $k_z=0$ plane, as shown in Fig.~\ref{NpBi}(c). Due to the presence of $\{C_{3(111)}||C_{3(111)}|0,0,0\}$, the symmetry-related nodal lines appear in the $k_x=0$ and $k_y=0$ planes as well. The detailed coirrep analysis shows that the band inversion occurs between GM8- and GM7-coirrep bands at $\Gamma$, both of which are sixfold. The fourfold nodal lines are actually formed by this sixfold band inversion. All the generic bands are doubly degenerate 
due to the presence of two unitary SSG symmetries at an arbitrary $k$ point, $\mathcal{O}_\alpha=\{C_{2x}||E|0,\frac{1}{2},\frac{1}{2}\}$ and $\mathcal{O}_\beta=\{C_{2y}||E|\frac{1}{2},0,\frac{1}{2}\}$, with the anti-commutation relation  ($\{Q_\alpha,Q_\beta\}=0$). Once SOC is included, the symmetry is reduced and must be described by a MSG, in which the spin part and the lattice part are closely coupled. Thus, only some of the SSG operations, whose spin part and lattice part contain the same proper rotation, are preserved in the MSG (including SOC). As a subgroup of the SSG, the `msg.data' of MSG symmetry is automatically generated at the same time by the command `{\ttfamily irssg -ssg}'. One can obtain the MSG coirreps by the command `{\ttfamily irssg -nb 79 82 -so > irmsg.out}' (reading `msg.data' instead). This analysis is quite useful for the SOC-induced band splitting and other effects.
The results show that the SSG coirrep DT1 becomes the MSG coreps mDT1+mDT2, and DT2 also becomes mDT1+mDT2, as shown in Fig.~\ref{NpBi}(d) and (e). As a result, once SOC is included, the nodal line becomes fully gapped, since the crossing bands share the same MSG coreps.

\subsection{The \texorpdfstring{$E_{F}$}{EF}-enforced degenerate point in \texorpdfstring{Eu$_3$PbO}{Eu3PbO}}
\begin{figure}[t]
  \centering
  \begin{overpic}[width=0.3\linewidth]{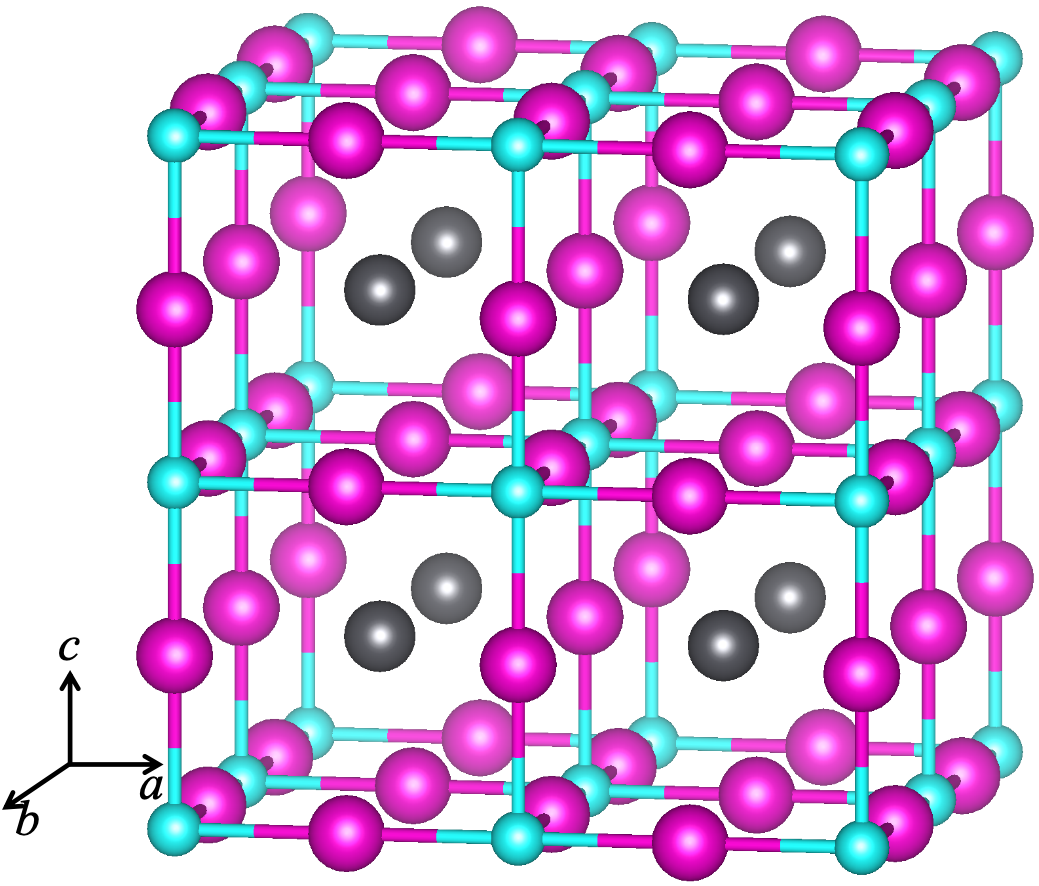}
    \put(-2,79){\small\textbf{(a)}}
  \end{overpic}%
  \begin{overpic}[width=0.3\linewidth]{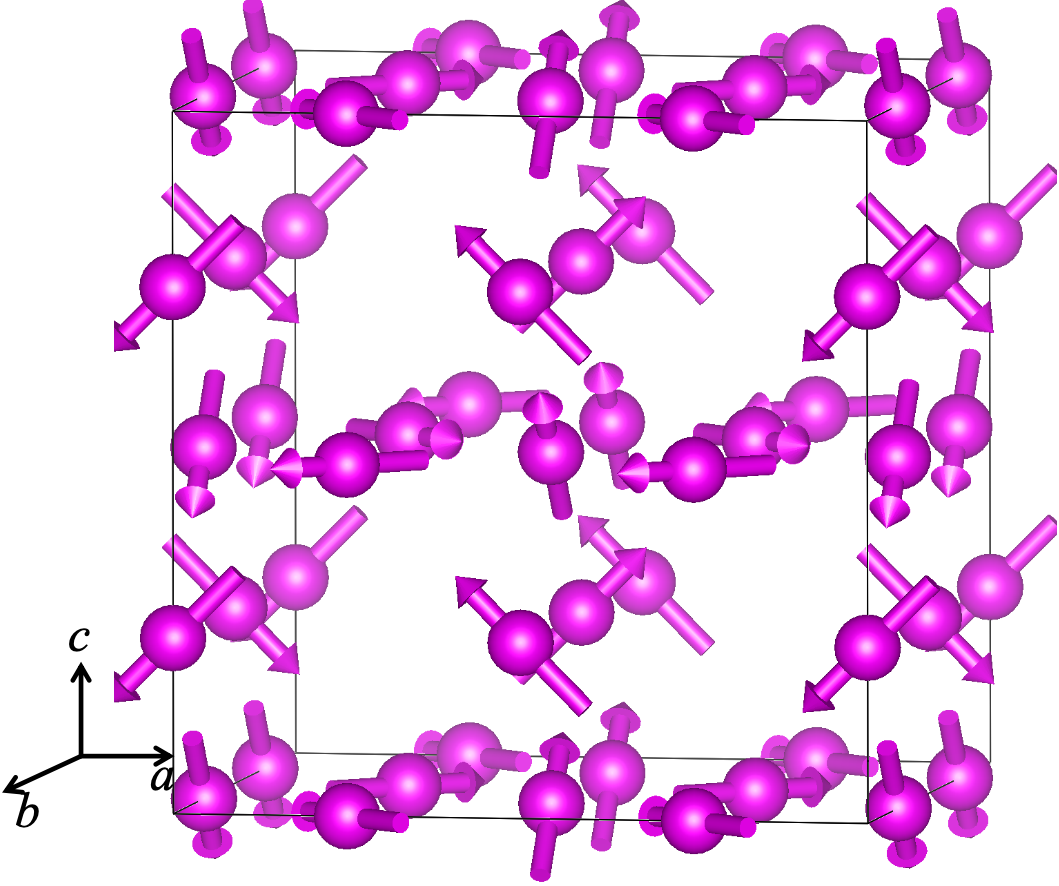}
    \put(0,79){\small\textbf{(b)}}
  \end{overpic}%
  \raisebox{-4mm}{%
    \begin{overpic}[width=0.38\linewidth]{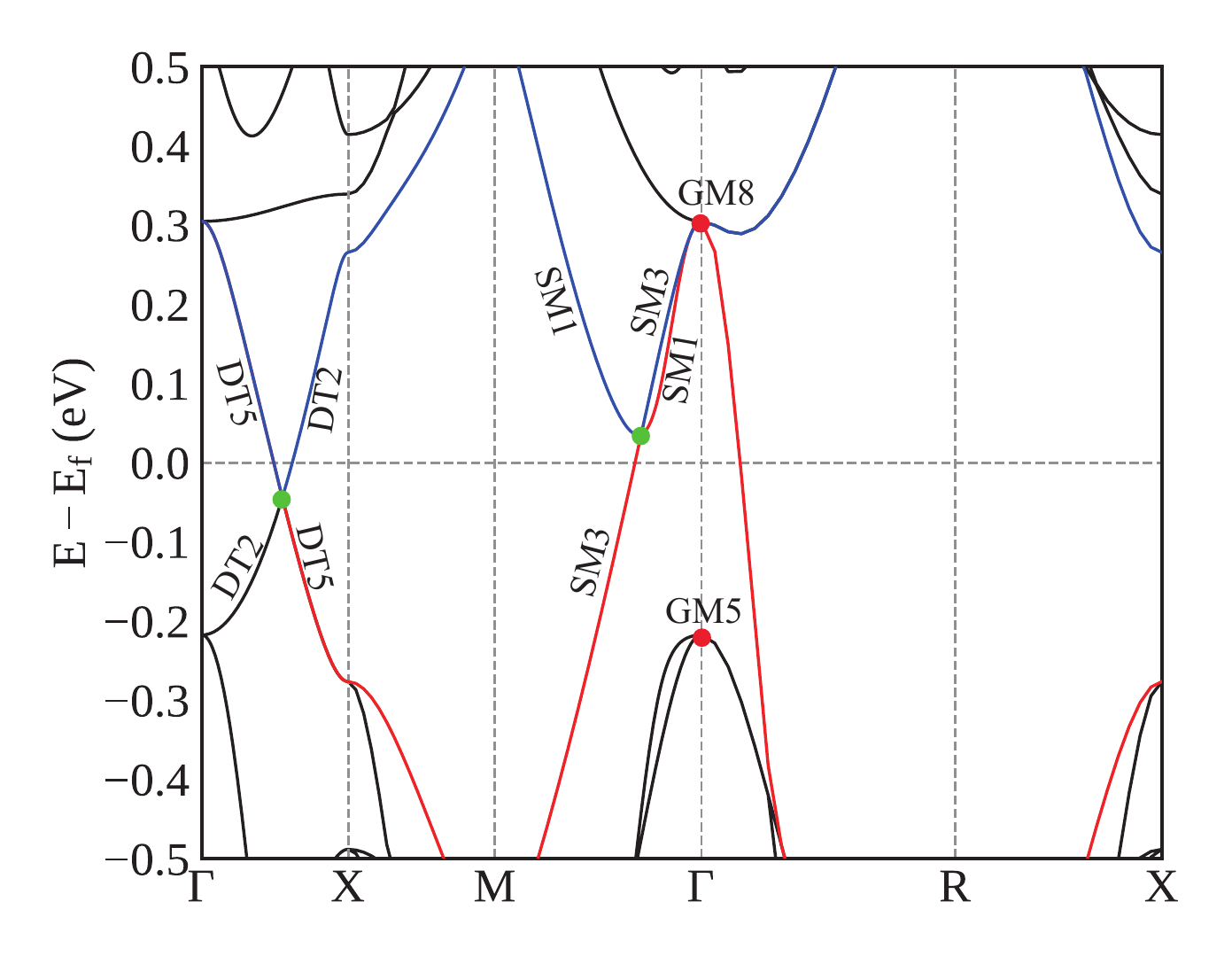}
      \put(0,68){\small\textbf{(c)}}
    \end{overpic}%
  }

  \caption{(a) Crystal structure of Eu$_3$PbO.
  (b) Magnetic structure of Eu in Eu$_3$PbO.
  (c) Magnetic energy bands of Eu$_3$PbO.}
  \label{EuPbO}
\end{figure}

The crystal structure of Eu$_3$PbO is shown in Fig.~\ref{EuPbO}(a), which is an anti-perovskite structure with Pb at the A site and O at the B site~\cite{PhysRevMaterials.6.114202}.
The noncoplanar magnetic configuration is illustrated in Fig.~\ref{EuPbO}(b), with SSG international symbol $P^{m,m,m} m^1 \bar{3}^3 m^m ({\rm O_h^{III}})$ and SSG number 221.8.6.1 obtained by {\ttfamily IRSSG}. After the noncollinear calculations in the absence of SOC, 
the band structure and the wavefunctions are obtained along high-symmetry lines (as shown in Fig.~\ref{EuPbO}(c)). Using the command `{\ttfamily irssg -nb 487 492 > irssg.out}', the coirreps 
are generated for bands 487-492 ($N_{tot}=488$). We conclude that the crossing along $\Gamma$X is formed by DT2 (twofold) and DT5 (fourfold). 
The detailed analysis shows that the band inversion occurs between GM5 (fourfold) and GM8 (sixfold), resulting in a high-degeneracy point between $N_{tot}$ and $N_{tot}+1$ bands. The red line is the highest valence band, while the blue line is the lowest conduction band. All the bands are doubly degenerate 
due to the presence of two unitary SSG symmetries with the anti-commutation relation ($\{C_{2x}||E|\frac{1}{2},\frac{1}{2},0\}$ and $\{C_{2y}||E|0,\frac{1}{2},\frac{1}{2}\}$).



\section{Conclusions}
\label{Conclusions}
In summary, we have developed an open-source software package -- {\ttfamily IRSSG} -- representing the first computational code to determine the coirreps of magnetic states in spin space groups.
It first generates the SSG operations, numbers, and international symbols for magnetic crystal structures; second, it constructs character tables for high-symmetry $k$ points; and finally, it computes the traces of SSG operations and assigns coirreps of the magnetic energy bands obtained from DFT codes. Detailed information is released on the HSP. In addition, it also works for the MSG symmetry once the SOC is included in the SSG systems. 
We further demonstrate how to use it to identify coirreps and high-degeneracy excitations in magnetic materials. 
This program does not restrict the user to a single DFT program or a single TB Hamiltonian. It works for the DFT codes, such as VASP and Quantum ESPRESSO, as well as any other code that has an interface to Wannier90. The program is very important for advancing research on the study of magnons, altermagnetism, magnetic topology, and novel high-degeneracy excitations in SSG systems.

\ \\
\noindent \textbf{Acknowledgment}

We thank Prof. Chen Fang and Dr. Yi Jiang for helpful discussions. This work was supported by the National Key R\&D Program of China (Grant No. 2022YFA1403800), National Natural Science Foundation of China (Grant No. 12188101), and the Center for Materials Genome.

\beginsupplement{}

\appendix
\section{Comparison between different SSG notations}
\label{appendix:ssg_notation_compare}

We provide a side-by-side comparison between our symbols output by {\ttfamily IRSSG} and Chen-Liu symbols of Ref.~\cite{PhysRevX.14.031038}, as shown in Table \ref{tab:ssg_symbol_compare_examples_rebuttal}. Chen-Liu symbols can be rather involved, as they are constructed within a classification scheme that divides SSGs into three types: $t$-type, $k$-type, and $g$-type. We use the three materials discussed in the main text (Mn$_3$Sn, NpBi, and Eu$_3$PbO), and additionally include CeAuAl$_3$ in Ref.~\cite{PhysRevX.14.031038}.

\begin{table}[h]
\centering
\begin{tabular}{lll}
\hline
Materials & Chen-Liu symbol of Ref.~\cite{PhysRevX.14.031038} & {\ttfamily IRSSG} symbol \\
\hline
Mn$_3$Sn &
$t$-type: $P^{3^{1}_{001}}\,6_{3}/\,{}^{1}m^{2_{100}}\,m^{2_{\pi/3}}\,c^{m}\,1$ &
$P^{1,1,1} 6_3^{3^{-1}} / m^1 m^m c^m (\mathrm{C_{3v}^{II}})$\\
NpBi &
$g$-type: $F^{1}\, m^{3^{1}_{111}}\bar{3}^{m_{1\bar{1}0}}\, m \mid (1,1,1;\,2_{001},\,2_{010},\,2_{100})$ &
$F^{2,2,2} m^1 \bar{3}^{3^{-1}} m^m ({\rm T_d^{III}})$\\
Eu$_3$PbO &
$g$-type: $P^{1} m^{3^{1}_{111}} \bar{3}^{m_{01\bar{1}}}\, m \mid (m_{010},\, m_{001},\, m_{100})$ &
$P^{m,m,m} m^1 \bar{3}^3 m^m ({\rm O_h^{III}})$ \\
CeAuAl$_3$ & $k$-type: $P^{1}4^{1}m^{1}m^{{4}^1_{001}}\left(\frac{1}{2}\ \frac{1}{2}\ \frac{1}{4}\right)^{m}{1}$ & $I^{4^{-1}4^{-1}4} 4^1 m^1 m^1 (\rm{C_4^{II}})$\\
\hline
\label{irssgvsfindspingroup}
\end{tabular}
\caption{Several magnetic materials, with their SSG symmetries represented in Chen-Liu symbol and in our symbol by {\ttfamily IRSSG}.
}
\label{tab:ssg_symbol_compare_examples_rebuttal}
\end{table}
\section{Derivation of the factor system arising from the translation part}\label{appendix:proof_factor}
Let ${\cal O_\alpha} = \{ U_\alpha|| R_\alpha|\bv_\alpha \}$, ${\cal O _\beta} = \{ U_\beta|| R_\beta|\bv_\beta \}$ and ${\cal O _\gamma} = \{U_\alpha U_\beta||R_\alpha R_\beta|(R_\alpha\bv_\beta + \bv_\alpha) \text{ mod }1\}$.
We aim to verify that the translation-derived factor takes the form
\begin{equation}
\omega_\tau({\cal O}_\alpha,{\cal O}_\beta) = \exp\!\left[-i \bk\cdot (R_\alpha-\det(U_\alpha))\bv_\beta\right],
\label{omega_t}
\end{equation}
and to show how this factor arises from the relation between linear and projective representations.

We relate the projective representation $M$ and the linear representation $D$ by a translational phase factor:
\begin{equation}
    D({\cal O_\alpha}) = \exp(  -i\bk\cdot \bv_\alpha) M({\cal O_\alpha}).
    \label{projective and linear}
\end{equation}

Using the usual multiplication law for the linear representations,
\begin{equation}
    D({\cal O_\alpha})D({\cal O_\beta})=D(\{U_\alpha U_\beta||R_\alpha R_\beta|R_\alpha\bv_\beta + \bv_\alpha\}),
    \label{linear mul}
\end{equation}
one expects the projective representations to satisfy a modified multiplication rule with a factor system,
\begin{equation}
    M({\cal O_\alpha})M({\cal O_\beta})=\omega_\tau({\cal O}_\alpha,{\cal O}_\beta)M({\cal O_\gamma}),
    \label{projective mul}
\end{equation}

We now derive $\omega_\tau$ by substituting Eq. \eqref{projective and linear} into the product $M({\cal O_\alpha})M({\cal O_\beta})$.
When ${\cal O_\alpha}$ is unitary (so that $\det(U_\alpha) = 1$), a straightforward manipulation yields
\begin{equation}
    \begin{aligned}
    M({\cal O_\alpha})M({\cal O_\beta})&=\exp[i\bk\cdot (\bv_\alpha+\bv_\beta)]D({\cal O_\alpha})D({\cal O_\beta})\\
    &= \exp[i\bk\cdot (\bv_\alpha+\bv_\beta)]D({\cal O_\gamma})\exp[-i\bk\cdot(R_\alpha\bv_\beta+\bv_\alpha-\bv_\gamma)]\\
    &= \exp[i\bk\cdot (\bv_\alpha+\bv_\beta)]M({\cal O_\gamma})\exp[-i\bk\cdot \bv_\gamma]\exp[-i\bk\cdot(R_\alpha\bv_\beta+\bv_\alpha-\bv_\gamma)]\\
    & = \exp[i\bk\cdot (\bv_\beta - R_\alpha \bv_\beta)]M({\cal O_\gamma})
    \end{aligned}
    \label{alpha unitary}
\end{equation}
from which the unitary case of Eq. \eqref{omega_t} follows.

When ${\cal O_\alpha}$ is anti-unitary (so that $\det(U_\alpha) = -1$), one must account for the fact that both $M({\cal O_\alpha})$ and $D({\cal O_\alpha})$ reverse the sign of $i$, \ie $M({\cal O_\alpha})\,i = -i\, M({\cal O_\alpha})$ and $D({\cal O_\alpha})\,i = -i\, D({\cal O_\alpha})$. Repeating the same substitution and keeping track of this sign change leads to 

\begin{equation}
    \begin{aligned}
    M({\cal O_\alpha}) M({\cal O_\beta})&=\exp[i\bk\cdot (\bv_\alpha-\bv_\beta)]D({\cal O_\alpha}) D({\cal O_\beta})\\
    &= \exp[i\bk\cdot (\bv_\alpha-\bv_\beta)]\exp[-i\bk\cdot(R_\alpha\bv_\beta+\bv_\alpha-\bv_\gamma)]D({\cal O_\gamma})\\
    &= \exp[i\bk\cdot (\bv_\alpha-\bv_\beta)]\exp[-i\bk\cdot(R_\alpha\bv_\beta+\bv_\alpha-\bv_\gamma)]\exp[-i\bk \cdot \bv_\gamma]M({\cal O_\gamma})\\
    & = \exp[-i\bk\cdot (\bv_\beta + R_\alpha \bv_\beta)]M({\cal O_\gamma})
    \end{aligned}
    \label{alpha antiunitary}
\end{equation}
This yields the anti-unitary case of Eq. \eqref{omega_t}.

Combining Eqs. \eqref{alpha unitary} and \eqref{alpha antiunitary}, we have thus shown that any linear representation $D({\cal O_\alpha})$ obeying Eq. \eqref{linear mul} can be obtained from a projective representation $M({\cal O_\alpha})$ that satisfies Eq. \eqref{projective mul} with the factor system given in Eq. \eqref{omega_t}. The two representations are related by the translational phase factor in Eq. \eqref{projective and linear}.

\section{`tbbox.in' parameters}\label{tbbox}
For tight-binding calculations, \texttt{tbbox.in} is required. A useful tool `gen\_tbbox.py' is developed to generate \texttt{tbbox.in} from Wannier90 files directly, which is available at \href{https://github.com/zjwang11/IRSSG}{https://github.com/zjwang11/IRSSG}. An example of Mn$_3$Sn with a type-II configuration is given below.
\begin{lstlisting}
  spinpol = False 
  #hr_name_up = sp.up_hr.dat 
  #hr_name_dn = sp.dn_hr.dat
  hr_name = symm_hr.dat

  
 proj:
 orbt = 2
 spincov = 1
 ntau = 8
  0.838800  0.677599  0.250000  2.598077  1.5  0  1 5  ! x1,x2,x3,m1,m2,m3,itau,iorbit
  0.161200  0.838800  0.750000 -2.598077  1.5  0  1 5
  0.838800  0.161200  0.250000 -2.598077  1.5  0  1 5
  0.161200  0.322401  0.750000  2.598077  1.5  0  1 5
  0.322401  0.161200  0.250000  0.000000 -3.0  0  1 5
  0.677599  0.838800  0.750000  0.000000 -3.0  0  1 5
  0.333333  0.666667  0.250000  0.000000  0.0  0  2 3
  0.666667  0.333333  0.750000  0.000000  0.0  0  2 3
 end proj
 
 kpoint:
 kmesh = 10
 Nk = 2
  0 0 0
 0.5 0 0
 end kpoint
 
 unit_cell:
     5.665000    0.000000    0.000000
    -2.832500    4.906034    0.000000
     0.000000    0.000000    4.531000
 end unit_cell

\end{lstlisting}

\begin{enumerate}
\item \texttt{spinpol}: whether the magnetism of the system is collinear (type-I SSG) (\eg  `ISPIN=2' and `NONCOLLINEAR=.FALSE. in \texttt{INCAR} for VASP'). Default: \texttt{False}.
\item \texttt{hr\_name}: the name of the input file containing TB parameters, whose format should be the same as `wannier90\_hr.dat' generated by Wannier90. This parameter only works when \texttt{spinpol=False}.
\item \texttt{hr\_name\_up}: the name of the input file containing TB parameters with up spin. This parameter is valid for \texttt{spinpol=True}.
\item \texttt{hr\_name\_dn}: the name of the input file containing TB parameters with down spin. This parameter is valid for \texttt{spinpol=True}.
\item \texttt{unit\_cell}: $3\times3$ lattice matrix in Cartesian coordinates (Å). Each row is one direct lattice vector, in the order $\mathbf a$, $\mathbf b$, $\mathbf c$. The 3-5 lines in \texttt{POSCAR} can be pasted here directly.
\item \texttt{kpoint} … \texttt{end kpoint}:
  defines a k-space path for band sampling.
  \begin{itemize}
    \item \texttt{kmesh}: number of evenly spaced samples taken on each consecutive line segment between listed k-nodes.
    \item \texttt{Nk}: number of $k$ points that follow; the path is formed by connecting them in order.
    \item $k$ points list: $k$ points given in fractional coordinates with respect to the reciprocal basis vectors $\mathbf b_1,\mathbf b_2,\mathbf b_3$.
  \end{itemize}
\item \texttt{proj:} … \texttt{end proj}: defines the local-orbital projectors used to build the TB basis.
      Each projector line has eight fields \texttt{(x1, x2, x3, m1, m2, m3, itau, iorbit)}.
  \begin{itemize}
    \item \texttt{orbt}: convention for orbital ordering. It selects how each \texttt{iorbit} maps to concrete orbitals, as in Table~\ref{tab:orbt_conv1}.
    \item \texttt{spincov}: \emph{spinful basis ordering}. 
          \texttt{1} = orbit-major, spin-minor (\(a\uparrow,\, b\uparrow,\,\dots,\, a\downarrow,\, b\downarrow,\,\dots\)); 
          \texttt{2} = spin-major, orbit-minor (\(a\uparrow,\, a\downarrow,\, b\uparrow,\, b\downarrow,\,\dots\)). 
    \item \texttt{ntau}: number of site lines that follow (i.e., the number of TB sites listed in this block).
    \item Site lines: each line is \texttt{(x1, x2, x3, m1, m2, m3, itau, iorbit)} where 
          \texttt{x1, x2, x3} are fractional coordinates, 
          \texttt{m1, m2, m3} are magnetic moments in Cartesian coordinates, 
          \texttt{itau} labels the atomic species/type, and 
          \texttt{iorbit} selects the on-site orbital set according to \texttt{orbt} (see Table~\ref{tab:orbt_conv1}).
  \end{itemize}

\end{enumerate}

\begin{table}[htbp]
  \centering
  \caption{Local orbitals ordering in different conventions~\cite{GAO2021107760}.}
  \begin{ruledtabular}
  \begin{tabular}{cll}
    \textbf{iorbit} & \textbf{Convention 1} & \textbf{Convention 2} \\
    \hline
    1 & $s$ & $s$ \\
    3 & $p_x,\,p_y,\,p_z$ & $p_z,\,p_x,\,p_y$ \\
    5 & $d_{xy},\,d_{yz},\,d_{xz},\,d_{x^{2}-y^{2}},\,d_{z^{2}}$ & $d_{z^{2}},\,d_{xz},\,d_{yz},\,d_{x^{2}-y^{2}},\,d_{xy}$ \\
    4 & $s,\,p_x,\,p_y,\,p_z$ & $s,\,p_z,\,p_x,\,p_y$ \\
    6 & $s,\,d_{xy},\,d_{yz},\,d_{xz},\,d_{x^{2}-y^{2}},\,d_{z^{2}}$ & $s,\,d_{z^{2}},\,d_{xz},\,d_{yz},\,d_{x^{2}-y^{2}},\,d_{xy}$ \\
    8 & $p_x,\,p_y,\,p_z,\,d_{xy},\,d_{yz},\,d_{xz},\,d_{x^{2}-y^{2}},\,d_{z^{2}}$ & $p_z,\,p_x,\,p_y,\,d_{z^{2}},\,d_{xz},\,d_{yz},\,d_{x^{2}-y^{2}},\,d_{xy}$ \\
    9 & $s,\,p_x,\,p_y,\,p_z,\,d_{xy},\,d_{yz},\,d_{xz},\,d_{x^{2}-y^{2}},\,d_{z^{2}}$ & $s,\,p_z,\,p_x,\,p_y,\,d_{z^{2}},\,d_{xz},\,d_{yz},\,d_{x^{2}-y^{2}},\,d_{xy}$ \\
    7 & $f_{xyz},\,f_{5x^{3}-xr^{2}},\,f_{5y^{3}-yr^{2}},\,f_{5z^{3}-zr^{2}},\,f_{x(y^{2}-z^{2})},\,f_{y(z^{2}-x^{2})},\,f_{z(x^{2}-y^{2})}$ & $f_{z^3},\,f_{xz^{2}},\,f_{yz^{2}},\,f_{z(x^2-y^2)},\,f_{xyz},\,f_{x(x^2-3y^2)},\,f_{y(3x^{2}-y^{2})}$ \\
  \end{tabular}
  \end{ruledtabular}
  \label{tab:orbt_conv1}
\end{table}


\bibliography{main}
\end{document}